\documentclass[namedreferences]{solarphysics}

\usepackage[hyperref,optionalrh,natbib]{spr-sola-addons} 
\usepackage{graphicx}        
\usepackage{savesym}
\savesymbol{iint}
\savesymbol{iiint}
\usepackage{amsmath,amssymb}        
\usepackage{color}           
\usepackage{breakurl}        
\usepackage{lmodern}
\usepackage{booktabs}
\usepackage{rotating}
\usepackage{multirow}
\usepackage{longtable, lscape}

\begin{document}

\begin{article}

\begin{opening}

\title{Real-time flare detection in ground-based H$\alpha$ imaging at Kanzelh\"ohe Observatory}

\author{W.~\surname{P\"otzi}$^{1}$\sep
        A.M.~\surname{Veronig}$^{1,2}$\sep
        G.~\surname{Riegler}$^{3}$\sep
        U.~\surname{Amerstorfer}$^{2}$\sep
        T.~\surname{Pock}$^{3}$\sep
        M.~\surname{Temmer}$^{2}$\sep
        W.~\surname{Polanec}$^{1}$\sep
        D.J.~\surname{Baumgartner}$^{1}$
       }
\runningauthor{P\"otzi et al.}
\runningtitle{Real-time flare detection in ground-based H$\alpha$ imaging at Kanzelh\"ohe Observatory}

   \institute{$^{1}$ Kanzelh\"ohe Observatory for Solar and Environmantal Research, University of Graz, Austria \\
                     email: \url{werner.poetzi@uni-graz.at}\\ 
                     email: \url{astrid.veronig@uni-graz.at} \\
                     email: \url{wolfgang.polanec@uni-graz.at} \\
                     email: \url{dietmar.baumgartner@uni-graz.at} \\
                $^{2}$ Institute of Physics/IGAM, University of Graz, Austria 
                 Graz University of Technology \\
                 email: \url{ute.amerstorfer@uni-graz.at} \\
                     email: \url{manuela.temmer@uni-graz.at} \\
                $^{3}$ Institute for Computer Graphics and Vision, 
                 Graz University of Technology, Austria \\
                     email: \url{gernot.riegler@icg.tugraz.at} \\
                     email: \url{pock@icg.tugraz.at}\\
             }

\begin{abstract}
 Kanzelh\"ohe Observatory (KSO) regularly performs high-cadence full-disk 
 imaging of the solar chromosphere in the H$\alpha$ and Ca\,{\sc ii}\,K spectral 
 lines as well as the solar photosphere in white-light. In the frame of 
 ESA's Space Situational Awareness (SSA) programme, a new system for 
 real-time H$\alpha$ data provision and automatic flare detection was developed 
 at KSO. The data and events detected are published in near real-time at  ESA's 
 SSA Space Weather portal  (\url{http://swe.ssa.esa.int/web/guest/kso-federated}). 
 In this paper, we describe the H$\alpha$ instrument, the image recognition 
 algorithms developed, the implementation into the KSO H$\alpha$ observing 
 system and present the evaluation results of the real-time data provision and 
 flare detection for a period of five months. The H$\alpha$ data provision 
 worked in $99.96$\% of the images, with a mean time lag between image recording 
 and online provision of 4~s. Within the given criteria for the automatic image 
 recognition system (at least three H$\alpha$ images are needed for a positive 
 detection), all flares with an area $\ge$50 micro-hemispheres and located 
 within $60^\circ$ of the Sun's center that occurred during the KSO observing 
 times were detected, in total a number of 87 events.
 The automatically determined flare importance and brightness classes were 
 correct in $\sim$85\%. The mean flare positions in heliographic longitude and 
 latitude were correct within $\sim$1$^\circ$. The median of the absolute 
 differences for the flare start times and peak times from the automatic 
 detections in comparison to the official NOAA (and KSO) visual flare reports 
 were 3 min (1 min). 
\end{abstract}
\end{opening}


\section{Introduction}
    \label{Introduction} 
 
Solar flares are  sudden enhancements of radiation in localized regions on the 
Sun. The radiation enhancements are most prominent at short (EUV, X-rays) and long 
(radio) wavelengths, with only minor changes in the optical continuum emission. 
However, flares are well observed in strong absorption lines in the optical part 
of the spectrum, most prominently in the H$\alpha$ Balmer line of neutral 
hydrogen at $\lambda = 656.3$~nm. 
Flares typically occur within active regions of complex magnetic configuration
\cite[e.g.][]{Sammis_2000}. They 
are the result of an impulsive release of magnetic energy previously stored in 
non-potential coronal magnetic fields via flux emergence and surface flows. The 
released energy is converted into the acceleration of high-energy particles 
(\opencite{Wiegelmann_2014}), 
heating of the solar plasma and mass motions 
\cite[e.g., reviews by][]{Priest_2002,Benz_2008,Fletcher_2011}. 
Flares may or may not occur in association with coronal mass ejections (CMEs). 
However, the association rate is a strongly increasing function of the flare 
importance, and in the strongest and most geo-effective events 
typically both occur together (\opencite{Yashiro_2006}). 

CMEs, flares and solar energetic particles (SEPs), which are accelerated either 
promptly by the flare or by the interplanetary shock driven ahead of fast CMEs, 
are the main sources for severe space weather disturbances at Earth. 
CMEs are only very limited accessible to observations from ground, due to their 
faint appearance and the stray light in the Earth atmosphere. They are best tracked 
in white-light images recorded from coronagraphs on space-based observatories. 
Flares are regularly observed at X-ray and (E)UV wavelengths 
from satellites, but they are also well observed from ground-based observatories
in the H$\alpha$ spectral line. 

Besides regular visual detection, reporting and classification of solar H$\alpha$ 
flares by a network of observing stations distributed over the globe, and 
collection at NOAA's National Geophysical Data Center (NGDC), there are also 
recent efforts to develop automatic flare detection routines. The detection 
methods range from comparatively simple image recognition methods based on 
intensity variation derived from running difference images (\opencite{Piazzesi_2012}),
region-growing and edge-based techniques (\opencite{Veronig_2000, Caballero_2013})
to more complex algorithms using machine learning (\opencite{Qahwaji_2010}, 
\opencite{Ahmed_2013}) or support vector machine classifiers (\opencite{Qu_2003}).
These methods have been applied to space-borne image sequences in the EUV and 
soft X-ray range \cite[e.g,][]{Qahwaji_2010,Bonte_2013,Caballero_2013}, but 
also to ground-based H$\alpha$ filtergrams 
\cite[e.g.,][]{Veronig_2000,Henney_2011,Piazzesi_2012,Kirk_2013}.

\begin{figure}    
   \centerline{\includegraphics[width=1.0\textwidth,clip=0]{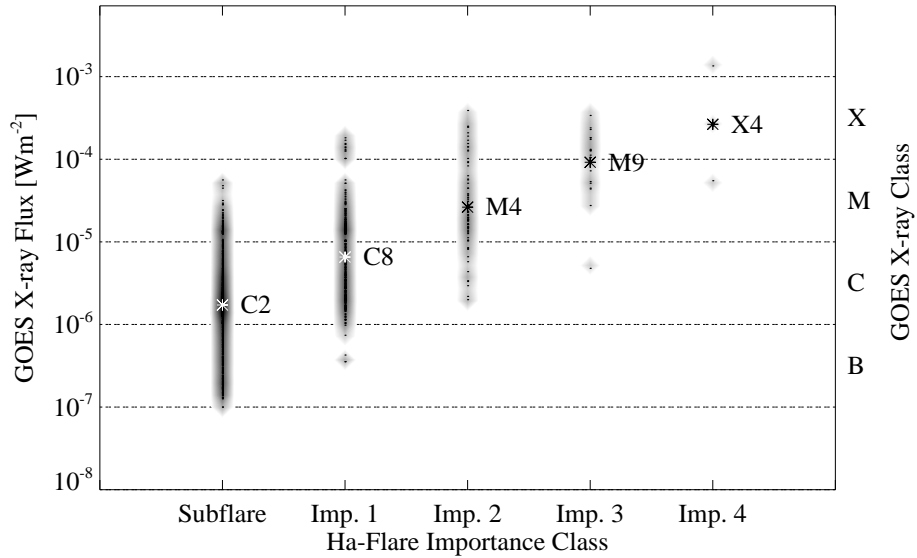}}
   \caption{GOES X-ray flares plotted against H$\alpha$ flares observed at KSO.
   The dark background represents the density of data points, the stars indicate 
   the mean of the logarithmic X-ray flare class}
   \label{xray_opt_fig}
 \end{figure}

\textbf{ The flare classification system used in this paper is based on the 
H$\alpha$ flare importance classification (\opencite{Svestka_1966}). 
Figure \ref{xray_opt_fig} shows the relation between the optical H$\alpha$ flare importance 
class and the X-ray flare class from the Geostationary Operational Environmental 
Satellites (GOES). The scatter plot contains all flares observed at KSO 
during the period 1/1975 - 10/2014 that were located within 60$\deg$ from the central 
meridian. The associated GOES X-ray flares were automatically identified by the 
following criteria: the soft X-ray and H$\alpha$ flare peak times are within 
10 min and the heliographic positions are within 10$\deg$. Figure  \ref{xray_opt_fig} 
reveals a high correlation between the H$\alpha$ importance class (defined by the chromospheric 
flare area; cf. Table \ref{tab:flares1}) and the GOES X-ray class (defined by 
the peak flux in the 1-8~{\AA} channel). In total the set comprises 2832 flares with the 
following distributions among the classes (H$\alpha$ importance: 81.2\% subflares, 
15.4\% importance 1, 2.6\% importance 2 and 0.8\% importance 3 and 4; GOES 
X-ray class: 86.0\% B and C, 12.5\% M and 1.5\% X-class flares).
} 



Space-based data have the advantage that there are 
no atmospheric disturbances (seeing, clouds) degrading the image quality, but 
there is  a delay in the data availability related to the data downlink. 
Ground-based data have the advantage that the data are immediately available 
for further processing, and can thus be efficiently used for the real-time 
detection and alerting of transient events such as solar flares in the frame 
of a space weather alerting system - however, with the drawback that 
the image sequences may suffer from data gaps and bad seeing conditions causing 
varying image quality. These circumstances have to be accounted for by the 
image recognition algorithms applied. 

In this paper, we present an automatic image recognition method that was developed 
for the real-time detection and classification of solar flares and filament eruptions
in ground-based H$\alpha$ imagery. The algorithms have been 
implemented into the H$\alpha$ observing system at Kanzelh\"ohe Observatory 
(KSO), in order to immediately process the recorded images and to provide 
the outcome in almost real-time.
This activity was performed in the frame of the space weather segment of 
ESA's Space Situational Awareness (SSA) programme, and the real-time 
H$\alpha$ data and detection results are provided online at
 \url{http://swe.ssa.esa.int/web/guest/kso-federated}. 
In this paper, we concentrate on the automatic flare detection and classification 
system, which was implemented in the KSO observing system in June 2013, and present 
the evaluation of the system for a five month period. The automatic detection of 
filaments and filament eruptions will be presented in a subsequent study, as the 
method is still under improvement \cite[first results are shown in][]{Poetzi_2014}. 

The paper is structured as follows. In Sect.~2, we describe the KSO solar 
instruments and observations. In Sect.~3, we outline the image recognition 
algorithms developed to automatically identify solar flares in  H$\alpha$ 
images, and to follow their evolution (in terms of location, size, intensity 
enhancement and classification). Sect.~4 outlines how the real-time detection 
and alerting was implemented in the KSO observing system. 
In Sect.~5, the outcome of the real-time 
flare detection system is evaluated for a test period of five months from 
end of June to November 2013. In Sect. 6, we discuss the performance of the 
system.


\section{KSO instrumentation and observations} 
      \label{Instrum}

\begin{figure}    
   \centerline{\includegraphics[width=1.0\textwidth,clip=0]{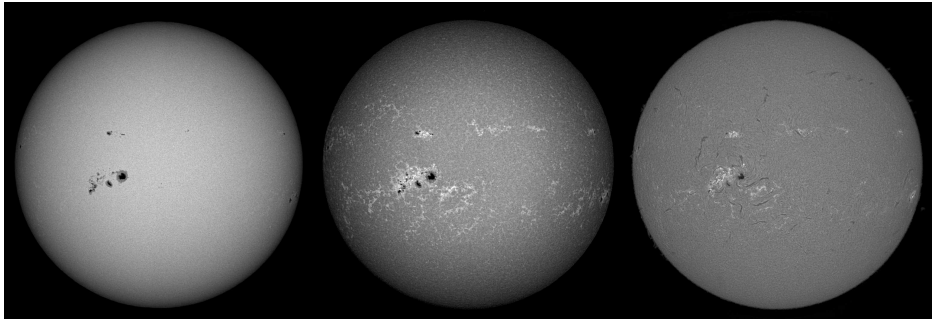}}
   \caption{Solar full-disk observations in white-light (left), Ca\,{\sc ii}\,K (middle) 
   and H$\alpha$ (right) recorded at Kanzelh\"ohe Observatory on January 6, 2014. }
   \label{sample_kso_fig}
 \end{figure}

Kanzelh\"ohe Observatory for Solar and Environmental Research 
(KSO; \url{http://kso.ac.at}) is operated throughout the year at a mountain ridge 
in southern Austria near Villach (N 46$^\circ$40.7', 13$^\circ$54.1', altitude 1526 m). 
The site allows solar observations for about 300 days a year, typically 1400 hours 
of observations. KSO regularly performs high-cadence full-disk observations of the Sun 
in the H$\alpha$ spectral line (\opencite{otruba_2003}), the Ca\,{\sc ii}\,K 
spectral line (\opencite{polanec_2011}), and in white-light 
(\opencite{otruba_2008}).  Figure~\ref{sample_kso_fig} shows an exemplary set 
of simultaneous KSO imagery in H$\alpha$, Ca\,{\sc ii}\,K and white-light 
recorded on January 6, 2014. All data are publicly available via the online KSO 
data archive at \url{http://kanzelhohe.uni-graz.at/} \citep{poetzi_2013a}.

\begin{figure}    
   \centerline{\includegraphics[width=1.0\textwidth,clip=0]{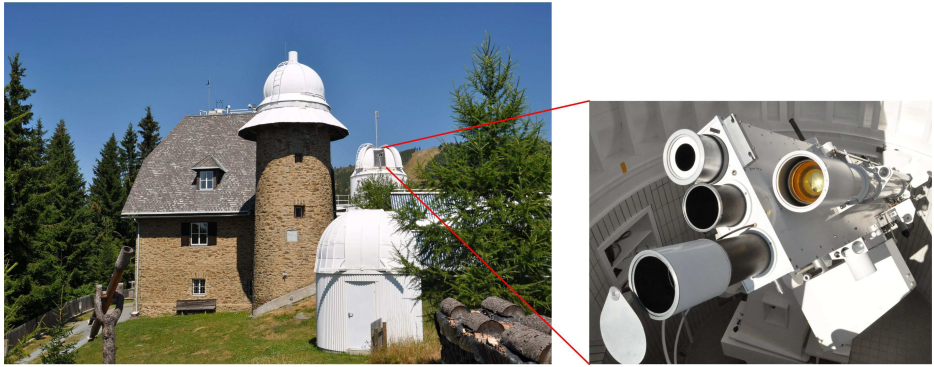}}
   \caption{Kanzelh\"ohe Observatory (left) and its solar patrol telescope 
   consisting of four refractors observing the Sun in H$\alpha$, Ca\,{\sc ii}\,K and 
   white-light (right).}
   \label{telescope_fig}
 \end{figure}

The observations are carried out by the KSO observing team during 7 days a week, 
basically from sunrise to sunset if the local weather conditions permit. 
All instruments for solar observations are mounted on the KSO surveillance 
telescope, which comprises four refractors on a common parallactic mounting 
(Figure \ref{telescope_fig}). 
The KSO H$\alpha$ telescope is a refractor with an aperture ratio number 
of d/f = 100/2000 and a Lyot band-pass filter centered at the H$\alpha$ spectral 
line ($\lambda = 656.3$\,nm) with a Full-Width-at-Half-Maximum (FWHM) of 0.07\,nm. 
For thermal protection an interference filter with an FWHM of 10\,nm
is in the light path. The Lyot filter can be tuned by turning the polarizers 
in narrow boundaries with little degradation of the filter characteristics. 
A beam splitter allows the application of two detectors at the same time. 
Currently, the observations are solely carried out in the center of the 
H$\alpha$ line. 

The CCD camera of the H$\alpha$ image acquisition system is a Pulnix TM-4200GE 
with 2048\,$\times$\,2048 Pixels and a Gigabit Ethernet interface. 
A frame rate of 7~images per second permits the application of frame selection 
(\opencite{roggemann_1994,Shine_1995}) to  benefit from moments of good seeing. 
The image depth of the CCD camera is 12 bit, which allows observing the quiet Sun and 
flares simultaneously without overexposing the 
flare regions. In order to have good counts statistics under varying seeing 
conditions and to avoid saturation effects in strong flares,
an automatic exposure control system is in place; the automatically controlled 
exposure time lies in the range 2.5 to 25\,milliseconds. 
In the standard observing mode,  the observing cadence of the H$\alpha$ telescope 
system is 6~seconds.
The plate scale of the full-disk observations is $\sim$1\,arcsec, corresponding 
to about 720~km on the Sun. The guiding of the telescope is performed by a 
microprocessor system, with (minor) corrections applied by automatically 
determining the solar disk center from the real-time H$\alpha$ images.

\section{Image recognition algorithm}
   \label{algorithm}

The developed image recognition algorithms make use of the main characteristics 
of the features in single H$\alpha$ images as well as in images sequences. 
Solar flares are characterized by a distinct brightness increase of localized 
areas on the Sun. They reach their maximum extent and maximum intensity typically 
within some minutes up to some tens of minutes, followed by a gradual decay of the 
intensity due to the subsequent cooling of the solar plasma. Flares are categorized in importance 
classes based on their total area and their brightness enhancement with regard 
to the quiet Sun level. 

The image recognition algorithm consists of four main building blocks.
The \textit{preprocessing} handles the different intensity distributions, 
large-scale inhomogeneities and noise. The \textit{feature extraction} step 
defines the characteristic attributes of the features to be detected, and 
how to model them. In the \textit{multi-label segmentation} step, the model 
is applied to ``new'', i.e. previously unseen, images in almost real-time. 
In the \textit{postprocessing}, every identified object is assigned and tracked 
via a unique ID, and the characteristic flare parameters are derived (location, 
area, start/peak time, etc.) In the following we give a basic description of 
these methods; further details can be found in \cite{Riegler_2013} and \cite{Riegler_2013b}.

\subsection{Preprocessing}
     \label{algo_pre}
     
The preprocessing has two goals, namely image normalization and feature 
enhancement. Across different H$\alpha$ image sequences, the intensity 
distributions of the images are shifted and dilated. These differences in the 
distributions arise due to different solar activity levels (e.g. many/few sunspots), 
seeing conditions, exposure time, etc. As the feature extraction strongly relies 
on value of the image intensities, we normalize the image intensities by a zero-mean
and whitening transformation:
\begin{align}
    \mu &= \frac{1}{|\Omega|} \sum_{x \in \Omega} f(x) \\
    \sigma &= \sqrt{ \frac{1}{|\Omega| - 1} \sum_{x \in \Omega} \left( f(x) - \mu \right)^2  } \\
    f_n(x) &= \frac{f(x) - \mu}{\sigma}
\end{align}
where $\Omega \subset \mathbb{R}^2$ is the image domain, $\mu$ the sample mean
and $\sigma$ the standard deviation of the input image $f$, respectively. 
The normalized image $f_n$ is given by a point-wise subtraction and division
by the mean and standard deviation.

As a second step in the preprocessing, additive noise and large-scale intensity 
variations, caused by the center-to-limb variation and clouds, are removed by 
applying a structural bandpass filter. At the core of this particular filtering method is the 
total variation with $\ell_1$ fitting term (TV-$\ell_1$) model (\opencite{Chan_2005, Aujol_2006}),
which is a signal and image denoising method  based on minimizing a convex optimization problem given by
\begin{align}
\min_u \| \nabla u \|_{2,1} + \lambda \| f - u \|_1
\end{align}
where $f$ is the noisy observation of the image and $u$ is the
minimizer of the optimization problem. The first term, the total
variation norm, regularizes the geometry of the solution and the
second term, the $\ell_1$ norm, ensures that the solution is close to
the original image $f$. Finally, $\lambda$ is a free parameter that
can be used to control the amount of regularization. The main property
of the TV-$\ell_1$ model is that it is contrast invariant. In other
words, structures from the image a removed only in terms of their
spatial extent and not in terms of their contrast to the background.
To efficiently solve the optimization scheme we use the generic primal
dual algorithm proposed in \cite{Chambolle_2011}. 

\begin{figure}    
\centerline{\includegraphics[width=0.99\textwidth,clip=]{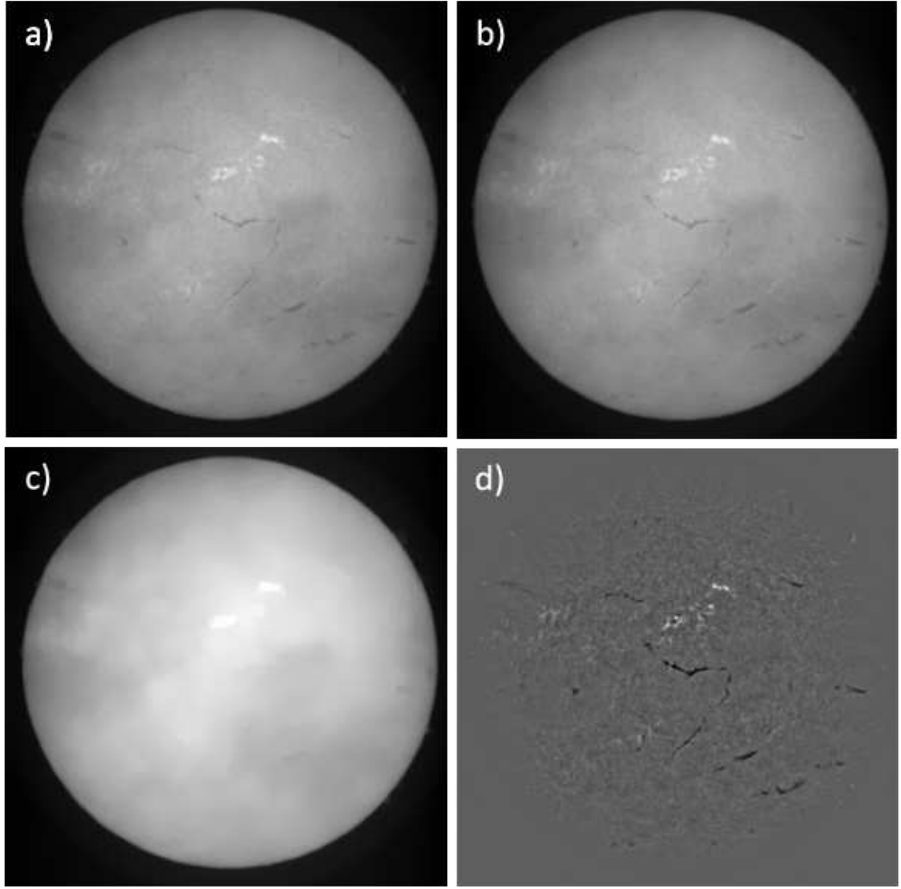}}
\caption{Structural bandpass filter applied to an H$\alpha$ image with clouds. 
a) Original image~$f$, 
b) denoised image $v_1$, c) large-scale variations $v_2$, 
d) resulting image $u$  of the structural bandpass filter.}
\label{bandpass_fig}
\end{figure}

It was shown in \cite{Chan_2005} that by solving the TV-$\ell_1$
for a certain parameter $\lambda$, all structures having a minimal
width of $\lambda^{-1}$ are removed in the regularized image $u$. We
utilize this fact in our structural bandpass filter by first removing
small-scale noise from the image using a larger $\lambda_1$, which
results in image $v_1$. In the next step, we remove larger structures
by again regularizing the image $v_1$ using a smaller $\lambda_2 <
\lambda_1$ such that the resulting image $v_s$ contains only unwanted
large scale structures such as brightness variations, clouds, etc.
The final result of the structural bandpass filter is then given by
subtracting image $v_2$ from image $v_1$.Figure~\ref{bandpass_fig} illustrates 
the different steps of the structural bandpass applied to a sample KSO H$\alpha$
filtergram with clouds.

 \subsection{Feature extraction}
     \label{algo_feat}

In the feature extraction step, two main problems have to be addressed. a) What 
are the characteristic attributes of flares and filaments, i.e. what
discriminates them from other solar regions? b) How can we efficiently model these attributes? 

To solve the first problem we assign to each pixel a feature vector.
The most intuitive feature choice is the pixel intensity of the
preprocessed images. We utilize the fact that filaments appear darker
than the background of the H$\alpha$ images, and that sunspots are even darker 
than filaments and have also different typical geometries
compared to filaments (round versus elongated objects). Flares are
defined as objects with distinctly higher intensities than the
background. It may also be useful to use the intensities of the pixels within a small
local neighborhood. 
\textbf{ Further, the contrast decreases from the center towards the limb. 
To incorporate this effect, the distance from the solar disk center to the pixel location
has proven to be useful.}

Based on the extracted feature vectors we utilize a Gaussian mixture model 
to assign each pixel of an H$\alpha$ image a class probability. 
For the classes we use the features ``flare", ``filament" and ``sunspot". 
The remaining part of the image is summarized in the class ``background".
 
\begin{figure}    
     \centerline{\includegraphics[width=01.0\textwidth,clip=]{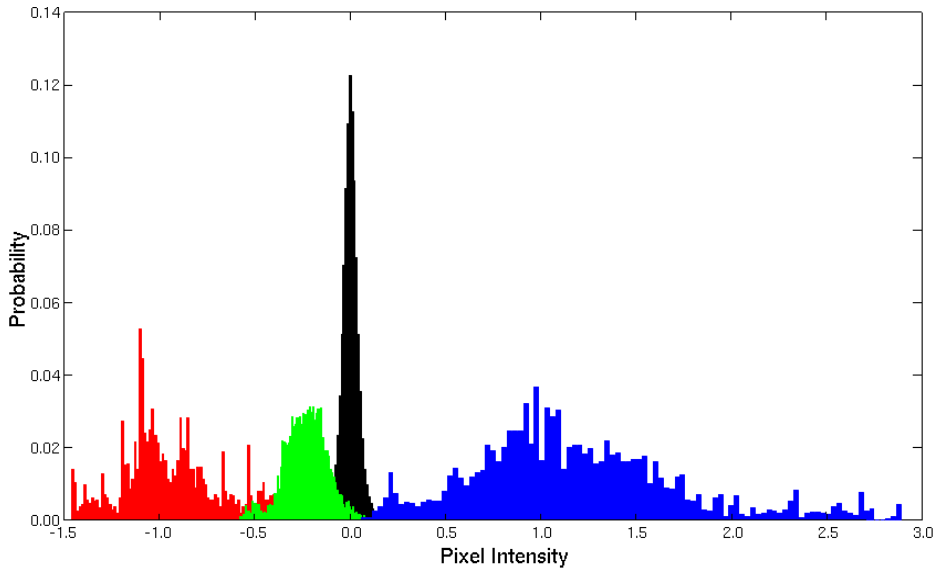}}
     \caption{Intensity distribution of the classes sunspot (red), filament (green), 
     background (black) and flare (red). The training examples are derived from 
     preprocessed H$\alpha$ images 
     that were annotated by an expert.  }
      \label{histogram_fig}
  \end{figure}

Figure~\ref{histogram_fig} illustrates the class probabilities in a histogram. 
The data that we use for the feature extraction and the learning of the model are 
derived from labeled H$\alpha$ images, where an expert annotated a set of KSO H$\alpha$  
images by assigning the pixels to the different classes. As one can see from the 
figure, the probability distributions of the four classes are not distinctly 
separated. The overlaps between the classes sunspot--filament and background--flare 
are no severe problem, because most of the probability mass is well separable. 
In contrast, the probability distribution overlap between the classes 
filament--background does cause segmentation problems in the application.  
Additional methods that can be used to arrive at a better distinction of filaments 
against the background are described in \cite{Riegler_2013b}. 
    
\subsection{Multi-label segmentation}
 \label{algo_multi}

In principle, each pixel could be assigned to the class with the highest probability, 
however, this would lead to a very noisy segmentation. In order to regularize the final
segmentation, we adopt a total variation based multi-label image segmentation model 
(\opencite{Chambolle_2011}):
\begin{align}
    \min_{ \{u_l\}_{l=1}^N } \sum_{l = 1}^N \int_\Omega \mathrm{d}|\nabla u_l| + 
    \sum_{l = 1}^N \int_\Omega u_l q_l \;\mathrm{d}x \\
    \text{s.t.} \quad u_l(x) \ge 0, \quad \sum_{l=1}^N u_l(x) = 1,
\end{align}
where the functions $u_l$ and $q_l$, $l=1...N$ are the binary class
assignment functions and class-dependent weighting function,
respectively. In the simplest case the negative logarithm of the class
probabilities can be used, but we apply an additional temporal
smoothing by computing an exponential weighted moving average over the
probabilities.
   
\begin{figure}    
   \centerline{\includegraphics[width=0.92\textwidth,clip=0,scale=0.8]{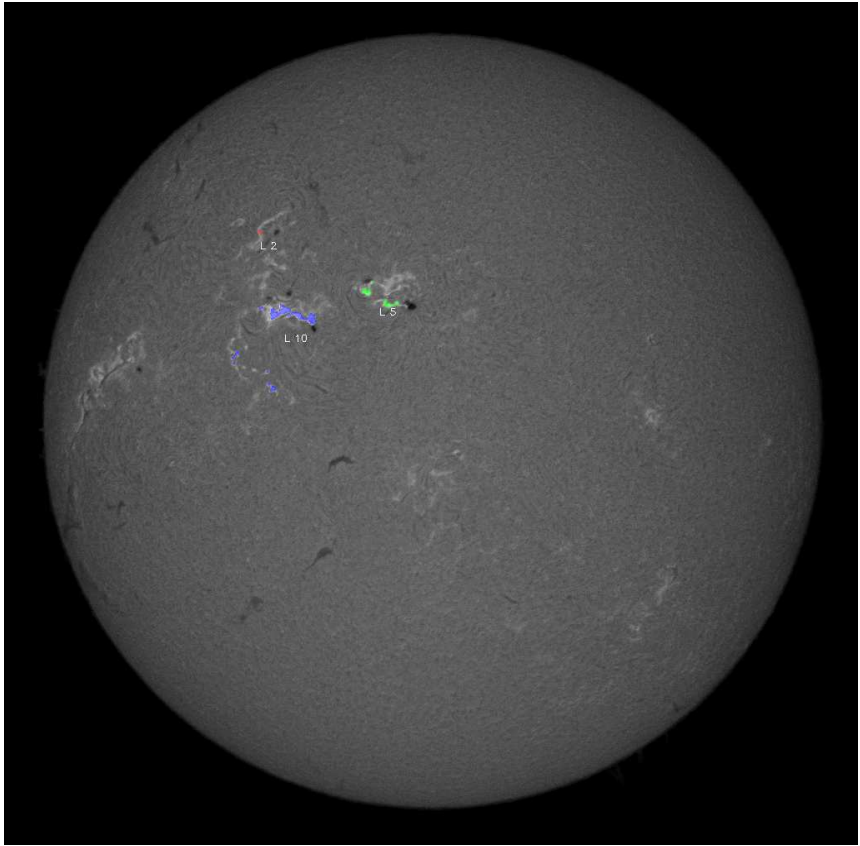}}
   \caption{Sample KSO H$\alpha$ image from May 10, 2014, in which three 
   different flares are simultaneously present on the solar disk. The detected 
   flare areas are indicated in different colors.
   Each flare is registered and tracked in time, i.e. from image to image by a unique ID. }
     \label{flares_fulldisk_fig}
 \end{figure}

\begin{figure}    
   \centerline{\includegraphics[width=0.92\textwidth,clip=]{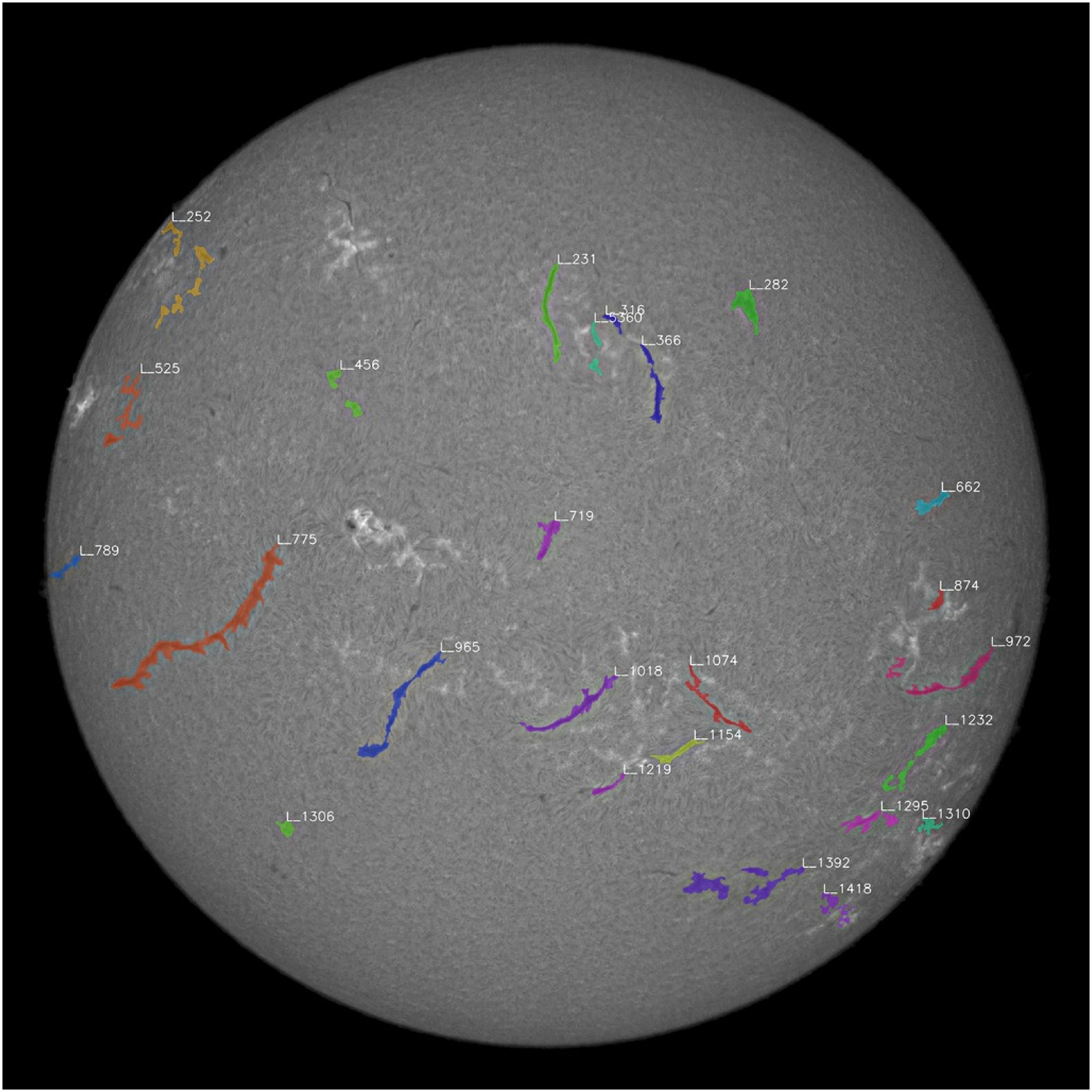}}
   \caption{Sample KSO H$\alpha$ image from March 31, 2014 together with the 
   filament detections. Each filament is assigned and tracked by its unique ID 
   (annotated at each filament). 
   }
   \label{filament_fig}
 \end{figure}

   \subsection{Postprocessing}
     \label{algo_post}

The final step of the method is the postprocessing that has two main goals. The 
first one is the identification of each detected flare (and filament) with a unique ID, 
which should remain the same over the image sequence for the very same object. 
The second goal is the derivation of characteristic properties from the identified 
objects to categorize them.

\subsubsection{Identification and tracking}

The identification task is solved by means of a connected-component
labeling problem (\opencite{Rosenfeld_1966}).  For the tracking of the
objects in the H$\alpha$ image sequence, we apply a simple propagation
technique. From the segmentation we obtain the four binary images
$u_l$ for the four classes. The next step is to identify 8-connected
pixels that form a group and are separated through zeros from other
groups, and to assign an ID to them.  The problem can be efficiently
solved with a two-pass algorithm as presented for example in
\cite{Haralock_1991}. In a first pass, temporary labels are assigned
and the label equivalences are stored in a union-find data
structure. Then a label equivalence is detected, whenever two temporary labels are
neighbors. In the second pass, the temporary labels are replaced by
the actual labels that are given by the root of the equivalence
class. The union-find data structure is a collection of disjoint sets
and has two important functions. The union function combines two sets,
and the find function returns the set that contains a given
number. The data structure can be efficiently implemented with trees.

The connected component labeling ensures that every flare (and filament) has 
a unique ID per image. To guarantee that the ID remains the same through the 
image sequence, we propagate the ID of previous images. Assume that $I_t (x)$ 
is the current component labeled segmentation and $\{I_{t-k}(x)\}_{k=1}^n$ the 
set of $n$ previous component labeled segmentation results. Then we change the 
ID of a current component $I_t$ to the ID $j$, where $j$ is given by the 
components of the previous images that have the most overlap with the 
component $I_t$. This can be implemented in a pixel-wise fashion and a simple 
map data structure. For a given component of the current image and ID $I_t$ we 
iterate all overlapping pixels $x$ of the set $\{I_{t-k}(x)\}_{k=1}^n$. 
If $I_{t-k} (x) \ne 0$,  we increment the counter for the ID $I_{t-k} (x)$ in 
the map. Finally, we assign the ID with the highest counter. 
Since flares often consist of two or more ribbons, flare detections that are 
located within a certain distance (set to 150 arcsec) are grouped to one ID. 
Figure~\ref{flares_fulldisk_fig} shows a sample H$\alpha$ image with the flare detections 
and the assigned flare IDs.
A sample H$\alpha$ image with the filament detections is shown in 
Figure~\ref{filament_fig}.


\section{\texorpdfstring{Implementation at the KSO H$\alpha$ observing system}{Implementation of the KSO H-alpha system}}
\label{data_proc}

To optimize the process and speed of the real-time data provision and flare detection,
different computers are involved that run in parallel, each one performing a specific set of tasks:
\begin{itemize}
    \item[--] camera computer: image acquisition; 
    \item[--] workstation 1: quality check, data processing and online data provision;
    \item[--] workstation 2: image recognition, flare detection and alerting.
\end{itemize}

 \begin{figure}
   \centerline{\includegraphics[width=1.0\textwidth,clip=]{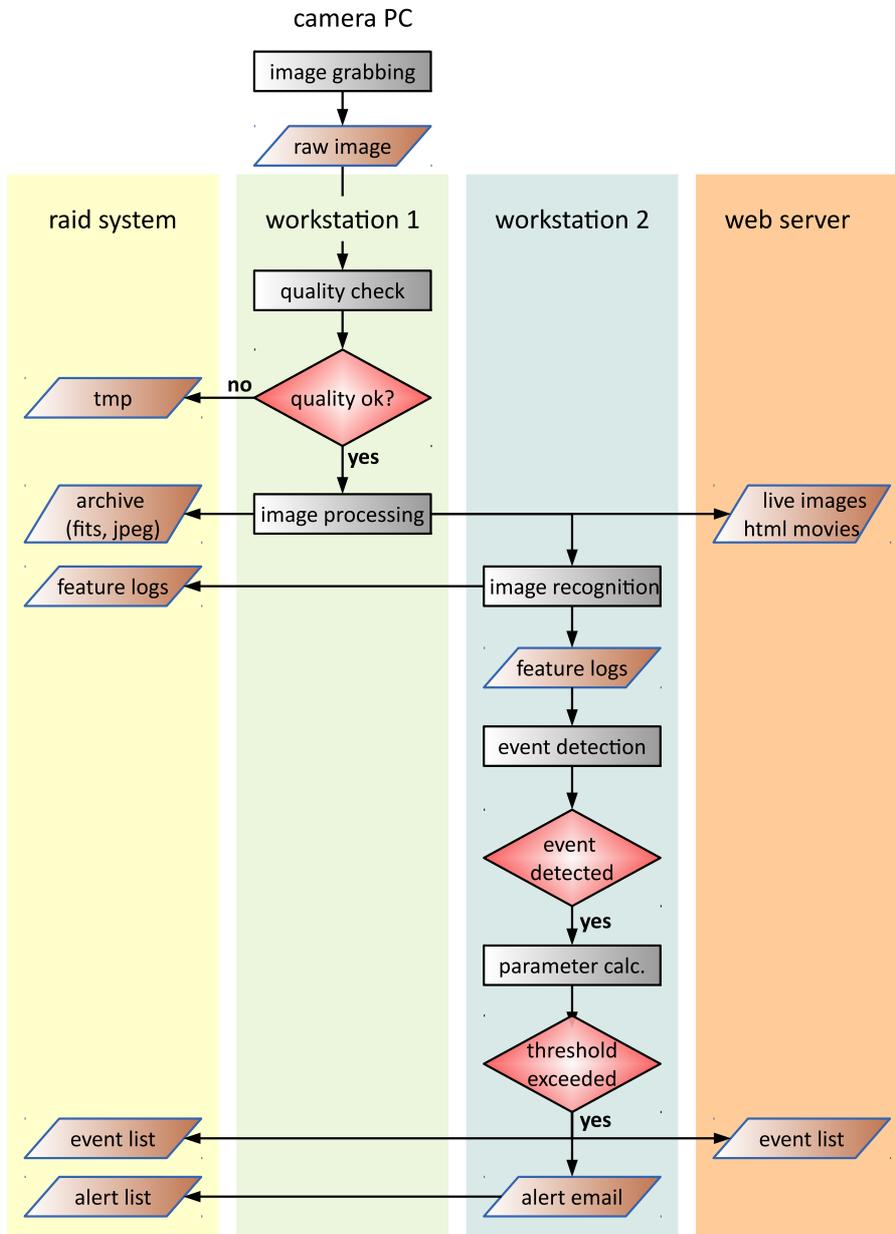}}
   \caption{Flow diagram showing the main steps of the data processing pipeline 
   at the KSO H$\alpha$ observing system. The overall process of data acquisition, 
   quality check, processing, image recognition, event detection and alerting 
   involves four different machines (camera PC, raid system, workstation 1 and 2) 
   and a web server where the results are published. 
   }
   \label{pipeline_fig}
 \end{figure}
 
Figure~\ref{pipeline_fig} shows a flow diagram of the tasks that are performed on 
each incoming H$\alpha$ image, where the columns refer to different machines 
responsible for certain tasks. Each H$\alpha$ image is grabbed by the camera 
computer and sent to workstation 1, where the image is checked for its quality. 
If the quality criterion is passed, the image is processed and published on the 
web server. In parallel, the processed image is also transferred to
workstation 2, on which the image recognition algorithm is performed. If an 
event is detected, its characteristic parameters are calculated. In case that 
the event exceeds a certain threshold (i.e. flare area/importance class), a 
flare alert is published online at ESA's SSA SWE portal and an alert email is 
sent out. In the following we give a detailed description of the different analysis steps.

 \subsection{Image acquisition, processing and online provision }
 \label{acquisition}

 \paragraph*{Image grabbing} The image acquisition is done in a fully automated mode, 
 which includes automatic  exposure control and the use of the frame selection 
 technique (\opencite{Shine_1995}). The CCD camera is controlled via a simple user 
 interface; in standard patrol mode no user interaction throughout the observation 
 day is needed. 
 
 \paragraph*{Quality check}
  
 All H$\alpha$ images grabbed are checked for their quality. Clouds and bad 
 seeing conditions result in low contrast and unsharp images, which may cause 
 difficulties for the image recognition.  Since the quality test has to be 
 performed on each image, a simple algorithm was implemented. The image quality 
 is measured by three conditions that have to be fulfilled:
 \begin{itemize}
    \item[--] The solar disk appears as a sphere with high accurateness:
    points on the solar limb are detected by a Sobel edge enhancement
    filter. A circle is fitted through the detected limb points and the
    relative error of the radius is computed.
    \item[--] The large-scale intensity distribution is uniform: the solar image 
    is rebinned to a 2\,x\,2 pixels image, and the relative brightness
    differences of these 4 pixels define a measure of the intensity distribution.
    \item[--] The image is sharp: the correlation between the raw image and a 
    smoothed version of the image is computed. If the raw image is already
    unsharp, it shows a high correlation with the smoothed image.
 \end{itemize}
 Based on these criteria, the images are classified in three quality groups: good, 
 fair and bad.  Only images of quality ``good" are sent to the image recognition 
 pipeline and  the online data provision. Images classified as ``fair" or ``bad" are 
 moved to a temporary archive and are not considered in the further analysis. 
 However, we note that images of quality ``fair" may still be acceptable and useful 
 for visual inspection (e.g. for visual flare detection).

 \paragraph*{Image processing}
 
 For all images that remain in the pipeline, decisive parameters like the disk center, 
 the solar radius, maximum and mean brightnesses, etc. are derived. Together with 
 additional information such as the acquisition time, instrument details, solar 
 ephemeris for the recording time, etc., the images are stored as FITS file 
 (\opencite{pence_2010}). These images are on the one hand stored in the KSO 
 data archive, on the other hand they are fed into the subsequent pipeline of 
 real-time data provision and image recognition.

 \paragraph*{Provision of real-time images and movies on the SSA SWE portal}
 
\begin{figure}
   \centerline{\includegraphics[width=1.0\textwidth,clip=]{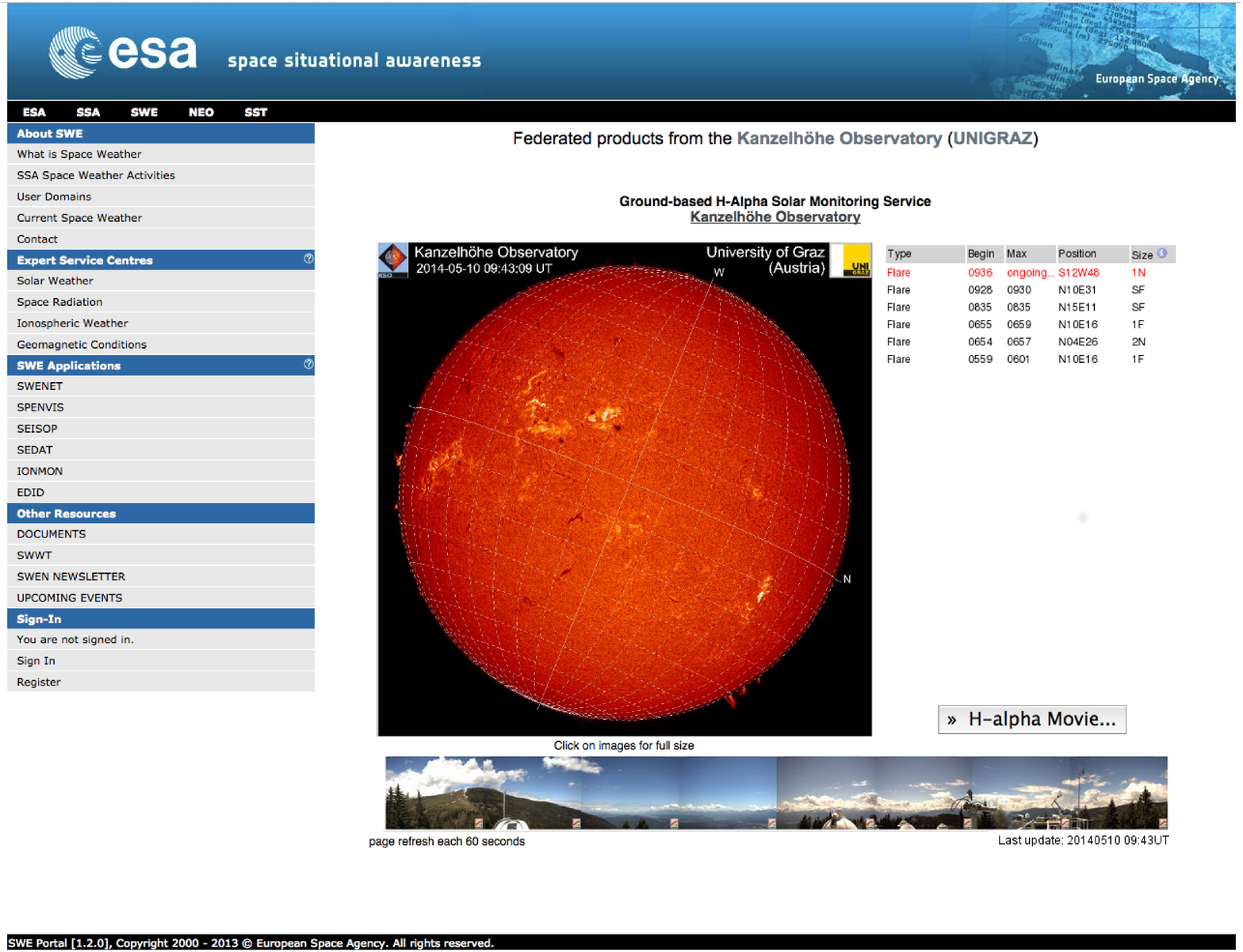}}
   \caption{Screenshot of the ESA SSA SWE H$\alpha$ subportal 
   (\url{http://swe.esa.int/web/guest/kso-federated}).
   The subportal shows the real-time H$\alpha$ image (middle), 
   a list of the detected events of that day (right) and a 360\,deg view 
   around the observatory (bottom) for checking the observating conditions, 
   and is updated every minute. }
   \label{portal_fig}
 \end{figure}

 Each minute an image is selected for the real-time H$\alpha$ display at 
 ESA's SSA SWE portal (\url{http://swe.ssa.esa.int/web/guest/kso-federated}; 
 a snapshot is shown in  Figure \ref{portal_fig}). The size of the image is reduced 
 to 1024$\times$1024 pixels and stored in jpeg format for fast and easy display. The 
 image is overlaid with a solar coordinate grid and annotated with a header
 containing time information. For later validation, with each image a log file is  
 updated that keeps track of the image acquisition
 time and the time when the image was provided online. 
 Every five minutes an html-movie script that animates the latest hour of 
 H$\alpha$ images is generated  and displayed at the SWE portal.

 \subsection{Image recognition, flare characterisation and alerting}
 
\paragraph*{Image recognition} 

Most of the iterative algorithms of the image recognition (cf.\ Sect.~\ref{algorithm}) 
are computationally intensive. However, they can be easily parallelized. Thus, it 
is possible to utilize the computational power of modern graphic processing units (GPU).
The image recognition algorithm has been implemented in the programming language 
C\raisebox{0.5ex}{\tiny\textbf{++}} and installed on a dedicated machine with a high-performance GPU. 
The system  benefits from the large number of processing units which are used for the 
highly parallelized computations. In its present form, the algorithm needs about 
10 s to  process the flare and filament recognition on one $2048 \times 2048$ 
pixels image, which allows event detection in near real-time. The results of the 
image recognition algorithm are stored in feature log files containing tables of 
flares and filaments that have been detected. These feature log files are updated at each 
time step, i.e. with each new image that enters the pipeline, so that the evolution 
of the detected features can be computed.

\paragraph*{Event detection and parameter calculation}

\begin{figure}    
   \centerline{\includegraphics[width=1.0\textwidth,clip=]{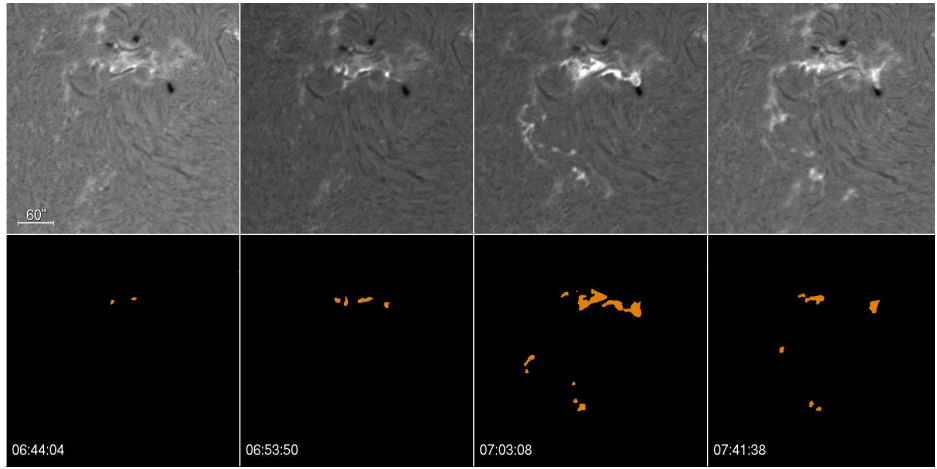}}
   \caption{Top: Sequence of H$\alpha$ images showing the evolution of a 2B 
   flare that occurred on May 10, 2014. Bottom: segmented flare areas. 
    }
   \label{flare_20140510_fig}
 \end{figure}

After the detection of a flare by the image recognition system, its characteristic 
properties and parameters are derived. For flares, these include the heliographic 
position, the flare area (which defines the importance class), the brightness 
class, and the flare start and peak times. These quantities need not only the 
information of a single H$\alpha$ image but also the information stored in the 
image recognition log files for the previous time steps. Handling of simultaneous 
flares is easily possible as each flare is identified via a unique ID that is 
propagated from image to image. In Figure~\ref{flare_20140510_fig} we show a 
sequence of H$\alpha$ images that were recorded during a 2B class flare that 
occurred on May 10, 2014 (top panels) together with the segmented flare 
regions (bottom panels).

The flare area is calculated by the number of segmented pixels with the same ID. These are 
subsequently converted by the pixel-to-arcsec scale of that day to derive the 
area in millionths of the solar hemisphere, so-called ``micro-hemisphere". 
The conversion procedure includes the information of the flare position to 
correct the effect of foreshortening toward the solar limb. The determined 
area is then directly converted to the flare importance class (subflares, 1, 2, 3, 4) 
according to the official flare importance definitions (cf. Table~\ref{flare_def}). 
For the categorization into the flare brightness classes (Brilliant-Normal-Faint: B-N-F), 
the intensity values relative to the background are used. To this aim, we compute 
the mean, standard deviation, maximum and minimum of the pixel intensities within 
the segmented regions. For each detected feature, we apply a normalisation by the 
difference between the maximum brightness and the mean brightness of the feature.  
 
  \begin{table}
   \caption{H$\alpha$ flare importance classes.}
    \label{flare_def}
    \begin{tabular}{ll}
      H$\alpha$ importance & Flare area \\
                                 & (micro-hemisphere) \\ \hline
      S[ubflares]                &  $<$ 100 \\
      1                          &  100 -- 250 \\
      2                          &  250 -- 600 \\
      3                          &  600 -- 1200 \\
      4                          &  $>$ 1200 \\ \hline & \\
    \end{tabular}
  \end{table}

\begin{figure}    
   \centerline{\includegraphics[width=1.0\textwidth,clip=]{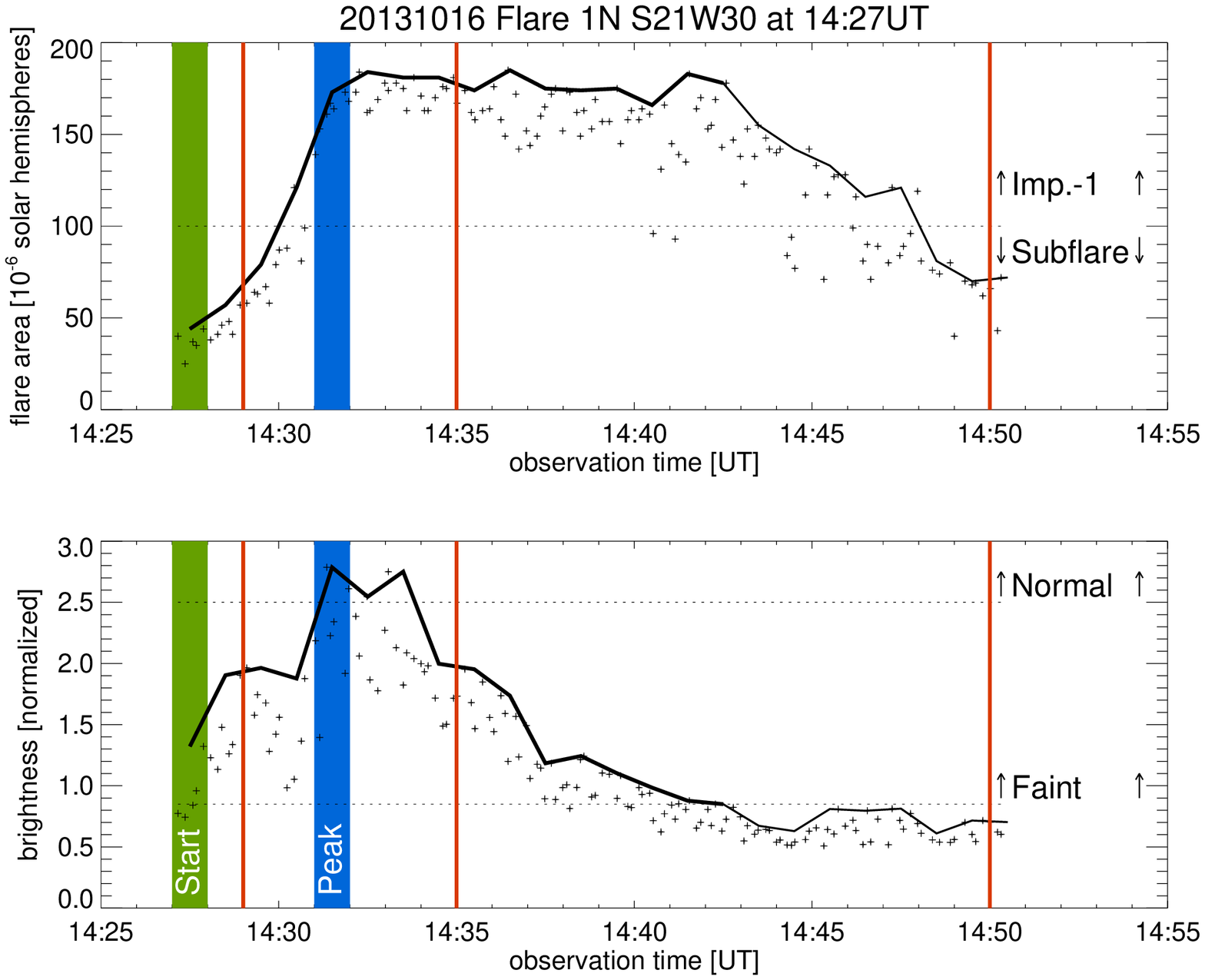}}
   \centerline{\includegraphics[width=1.0\textwidth,clip=]{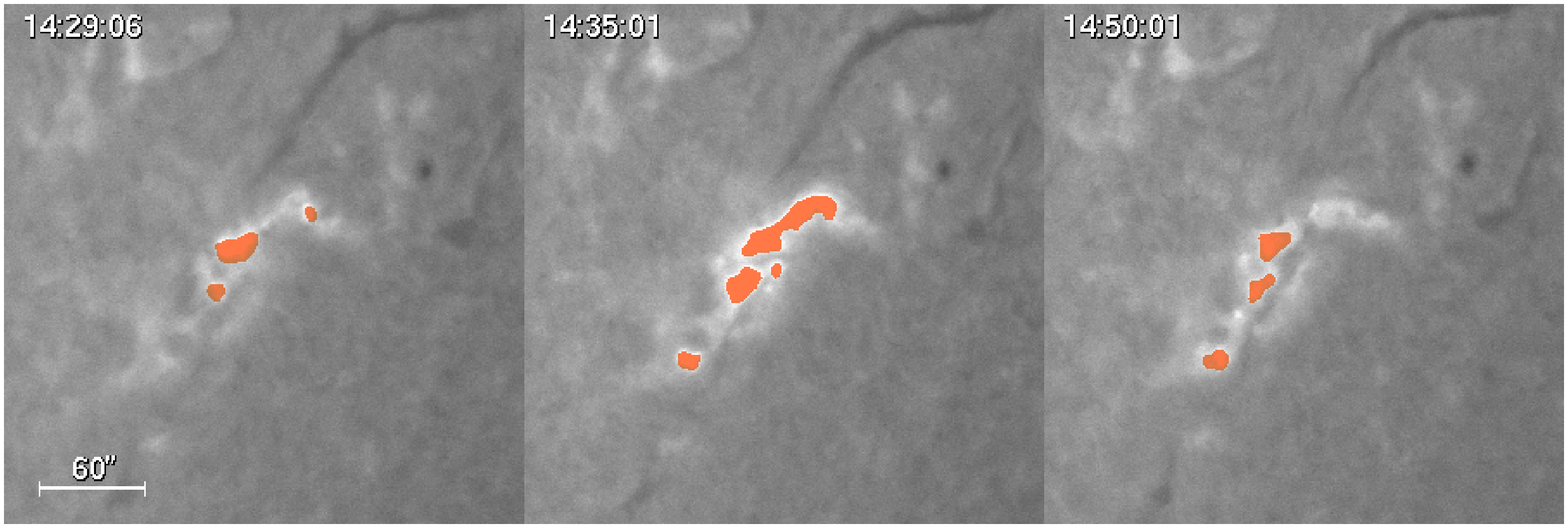}}
   \caption{Illustration of the flare parameter calculation for the 1N flare of 
   October 16, 2013.  Top: Evolution of the flare area. Middle: Evolution of 
   the flare brightness. The crosses show the data points, the 
   solid line shows the maximum values for each minute. The thick solid line represents
   times where the intensity is above the faint flare level. 
   The determined start (14:27 UT) and peak (14:31 UT) times are indicated by 
   the green and blue bar, respectively. Note that the maximum brightness and the
   maximum area do not necessarily occur at the same time.
   Bottom: Snapshots of the flare at three different times (indicated by 
   orange vertical lines on the top panels) together with the flare detections.
    }
   \label{area_bright_fig}
 \end{figure}

To characterize a flare, the evolution of the brightness and the area in each 
H$\alpha$ image of the sequence has to be analyzed. 
For illustration, we show in Figure~\ref{area_bright_fig} the evolution of the 
area and brightness of a sample 1N flare that occured 
on October 16, 2013. 
The flare classification is based on the following definitions:
 \begin{enumerate}
    \item The flare start is defined as the time when the brightness enhancement 
    is above the faint flare level for 3 consecutive images.
    \item The peak time of the flare is defined as the time where the maximum 
    flare brightness is reached. 
    \item The flare position is defined by the location of the brightest flare 
    pixel at the time of the flare peak.
    \item The importance class of the flare is defined via the maximum area of 
    the flare, and is updated when the area exceeds the level of a higher 
    importance class.
    \item The flare end is defined as the time when the brightness has decreased 
    below the faint level for 10 consecutive images or when there is a data gap 
    of more than 20 minutes. 
    \item  Handling of data gaps: In case of data gaps $<$20 min, the flare is 
    considered to be in an evolving state if the flare brightness after the 
    data gap is higher than before the gap.
    Data gaps of $>$20 minutes define the end of the flare.
  \end{enumerate}

\paragraph*{Flare alerts}

If a flare is detected that exceeds a certain threshold, i.e. importance class, 
then a flare alert is published on the ESA SSA SWE portal and an alert email is 
sent out to registered users. Originally, it was intended to restrict the 
flare alerting to events of H$\alpha$ importance class 1 and higher. However, 
due to the weak activity cycle 24, we lowered the threshold to subflares 
exceeding a size of 50 micro-hemispheres, in order to obtain a sufficient 
statistics for the evaluation.\footnote{In solar cycles 21 to 23 about 10\% of 
H$\alpha$ flares were larger than subflares \citep{Temmer_2001,Joshi_2005}. 
However, this number is much smaller in the current low-activity cycle 24. E.g. 
in the year 2013, only 35 of a total of 565 flares visually identified at KSO and 
reported to NOAA were larger than subflares.}

The event list on the ESA SSA SWE portal is updated every time  a flare is detected, 
when more information on flare becomes available during its evolution (e.g., the peak time) 
or when a flare that is already listed increases in its importance class. The flares 
are sorted in decreasing start time, so that the most recent event appears in the first 
line of the table. 
As long as the flare brightness increases, no peak time is listed but the event is 
annotated to be ``ongoing" and marked in red color in the event list in the H$\alpha$ 
subportal (cf.\ Fig.~\ref{portal_fig}).
When the flare brightness has decreased for $>$2 min, the peak time is derived and 
provided in the event table.
A list of all detected events is stored  in the local raid system for later evaluation.
 
In addition, flare alert emails are sent to a predefined list 
 of users. They are issued when one of the following criteria is fulfilled: 
 i) The flare detected is of importance class 1 or higher or a subflare with an area 
 exceeding 50 micro-hemispheres, or ii) an ongoing flare reaches a higher importance 
 class (e.g., a flare of importance 1 evolves further to a flare of importance 2).
 

\section{Results}

The system of near real-time H$\alpha$ data provision, automatic flare detection 
and alerting went online on June 26, 2013. In the following we present the results from 
evaluating the system for a period of five months from June 26 to November 30, 2013, 
in which it was run with the same set of parameters and definitions. 

\subsection{Real-time data provision}

To validate the online data provision, we evaluated the number 
of H$\alpha$ images that were recorded by the KSO observing programme and the 
number of H$\alpha$ images that were provided online in almost real-time at 
the ESA SWE service portal. For this purpose, log files recorded for each image 
both the observation time and the time when the image was put online to 
the SWE service portal. 

During the evaluation period, we had in total $563$ hours of solar observations 
at KSO. In total, 395\,129 H$\alpha$ images were recorded. $281\,806$ images (71.3\%)  
were rated as ``good", 20\,922 (5.3\%) as ``fair", and 92\,401 (23.4\%) as ``bad".
33\,765 H$\alpha$ images (one per minute) out of the ``good" sample  were provided 
online at the SWE service portal, whereas 14 were erroneously skipped due to internal 
data stream errors. This means that in $99.96$\% of the observation time, one 
image per minute was provided online at the SWE portal. The mean time lag 
between the recording of an image and its online provision was $3.6\pm 0.9$ seconds.

\subsection{Real-time flare detection and classification}

For the evaluation of the automated detection and alerting of H$\alpha$ flares, 
we considered all flares that exceeded an area of $50$ micro-hemispheres and that 
occurred within 60$^\circ$ from solar disk center during the KSO observing 
times.\footnote{For flares closer to the solar limb than 60$^\circ$ from the center of 
the disk, projection effects become significant in the determination of the 
flare area. In addition, these flares are most likely not relevant for space 
weather disturbances at Earth.} As discussed in Sect.~\ref{acquisition}, only 
images of quality ``good" are fed into the automatic flare detection pipeline. 
For the automatic detection  of a flare, we demand that it is observed in at 
least three H$\alpha$ images. Periods of $>$20~min containing no images of 
quality ``good" are defined as data gaps.

The data that are needed for the evaluation of the flare detection, classification 
and alerting are derived from the log files that are created and updated during 
the observations (cf.\ Fig.~\ref{pipeline_fig}). The relevant parameters that are 
derived and evaluated are:
\begin{itemize}
 \item[--] Flares: heliographic position, start time, peak time, area, 
     importance class, brightness class; 
 \item[--] Alerts: time of issue.
\end{itemize} 
For the evaluation, we compare the results obtained by the automated image 
recognition system developed, called Surya\footnote{Surya -``the Supreme Light"
is the chief solar deity in Hinduism}, against the official flare reports 
provided by the National Geophysical Data Center (NGDC) of the National Oceanic 
and Atmospheric Administration (NOAA) and by Kanzelh\"ohe Observatory. Both are obtained 
by visual inspection of the data by experienced observers. 

The Space Weather Prediction Center (SWPC) of the U.S. Department of Commerce, 
NOAA, is one of the national centers for environmental protection and provides 
official lists of solar events, online available at 
\url{http://www.swpc.noaa.gov/ftpmenu/indices/events.html}. 
The information on the flare events is collected from different observing stations 
from all over the world. Kanzelh\"ohe Observatory sends monthly flare reports 
to different institutions, including NGDC/ NOAA and the World Data Center (WDC) 
for Solar Activity (Observatoire de Meudon). The visual KSO flare reports (KSOv)
are online available at \url{http://cesar.kso.ac.at/flare_data/kh_flares\_query.php}.
We actually expect that the results of the automatic detections are on average 
closer to the visual KSO flare reports than the NOAA reports, as they 
are based on the data from the same observatory. 
However, it is also important to compare the outcome against the NOAA reports, 
as they provide an independent set of flare reports.

Table~\ref{tab:flares1} in the Appendix lists all flare events 
(area $\ge50$ micro-hemispheres; located within $60^{\circ}$ from disk center) which 
were detected in quasi real-time by the automated algorithm during the evaluation period. 
In total, $87$ flares were detected by Surya; $69$ were classified as subflares 
and $18$ as flares of importance 1. This list includes $3$ false detections 
(marked in red color in the table), i.e.\ flares that were detected by Surya but 
have no corresponding event reported by NOAA or KSOv. In addition, the list in 
Table~\ref{tab:flares1} includes 7 flares where Surya reported  one flare but 
NOAA and KSOv reported two separate events, as well as 2 NOAA (4 KSOv) flares, 
where Surya has split up one flare into two or more events.

To evaluate the detection ability of Surya, we checked also all NOAA and KSOv 
flares reported during the KSO observation times 
(located within 60$^\circ$ from disk center) that were not detected by Surya. 
These are in total $60$ flares (57 SF, 2 SN and one 1N flare). There are basically 
two different reasons why these events were not detected by Surya: 
\begin{enumerate}
 \item Data gaps: Less than 3 images of quality ``good'' were available during 
 the event, and thus the automatic detection algorithm was not run. 47 flares 
 fall into this category. We note that single images as well as images of lower 
 quality may still be sufficient to identify a flare by visual inspection 
 (though with large uncertainties in the derived flare parameters). Indeed, 
 for 19 out of these events visual flare reports from KSOv are available 
 (18 subflares, and one flare of importance~1).
 \item NOAA reports a flare which is not listed by KSOv and where -- even 
 after careful visual re-inspection of the KSO H$\alpha$ 
   image sequences -- we cannot confirm the appearance of a flare. 
   This applies to $13$ events, all of them subflares. 
\end{enumerate} 
Given the reasons above, these events are not expected to be detected by our 
automatic image recognition system. This means that Surya basically detected all 
flares listed by the NOAA and KSOv flare reports, when there were sufficient 
data available (i.e.\ at least three images during an event). 

The next question we have to address is how accurate are the flare parameters 
calculated by the automatic system. The accuracy of the peak flare area determined 
by Surya cannot be evaluated, since the flare reports do not provide areas. 
However, the flare area is an intrinsic property which determines the importance 
classification of a flare (cf. Table~\ref{flare_def}). For the importance 
classifications we find that there are 7 cases, in which the importance 
classes reported by NOAA and KSOv do not conincide. Thus we excluded 
those events from the evaluation of the importance classes as there is no 
unique reference value available. For the remaining 
set of flares, we find that in 86\% the automatically determined flare importance 
class coincides with the class given by NOAA and KSOv. The incorrectly classified 
events include 5 flares where Surya obtained a different importance class than 
reported by NOAA and KSOv, and 6 cases where Surya has split up flares in 2 or 
more events. For the brightness classification, there are 15 events where the 
NOAA and KSOv reports do not coincide. For the remaining set, we find that in 
85\% Surya determined the correct brightness class (cf.\ Table~\ref{tab:flares1}).

\begin{figure}
\centerline{
\includegraphics[width=1\textwidth,clip=]{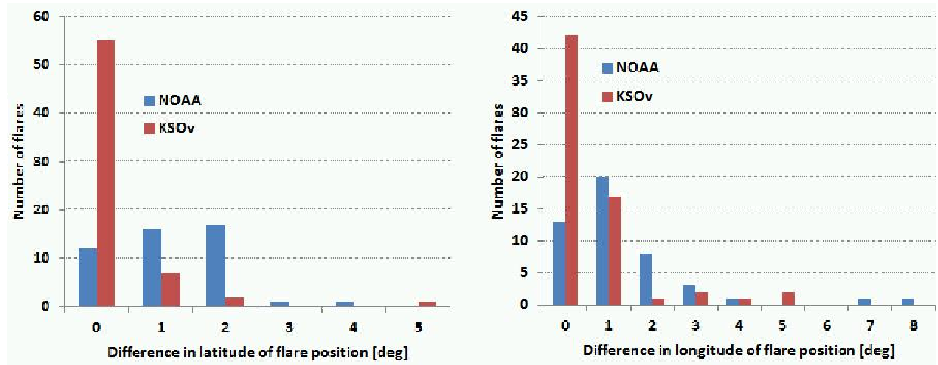}
}
   \caption{Distribution of the absolute differences of the flare's heliographic 
   latitude (left) and longitude (right) between the automatic detection by Surya 
   and the NOAA (blue) and KSOv (red) reports.}
   \label{fig:diffLon}
\end{figure}

Figure~\ref{fig:diffLon} shows the absolute differences of the heliographic 
latitude and longitude of the flare center as obtained by the automatic algorithm 
against the values reported by NOAA and KSOv. The mean of the absolute difference 
for the latitude is $1.21^{\circ}$ ($0.25^{\circ}$) with respect to the NOAA 
(KSOv) flare reports, and $1.36^{\circ}$ ($0.60^{\circ}$) for the longitude.

\begin{figure}
   \centerline{\includegraphics[width=0.95\textwidth,clip=]{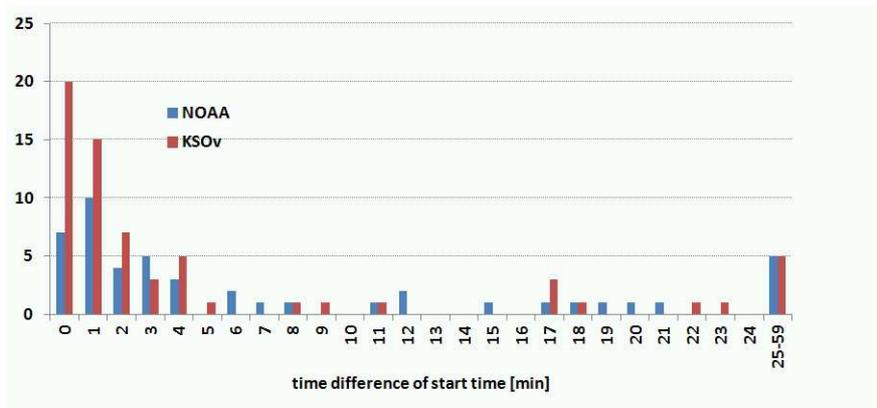}}
   \caption{Distribution of the absolute differences between start times of the flares 
   detected by Surya and reported by NOAA (blue) and KSOv (red).}
   \label{fig:diffStart}
\end{figure}

\begin{figure}
\centerline{\includegraphics[width=0.95\textwidth,clip=]{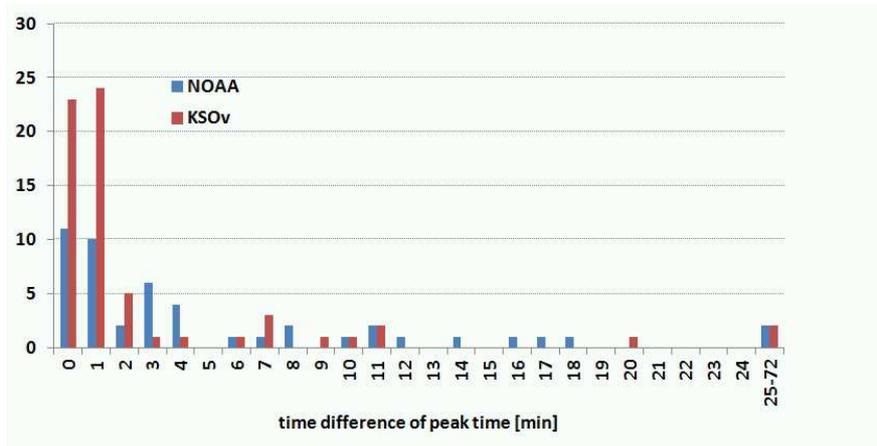}}
   \caption{Distribution of the absolute differences between peak times of the 
   flares detected by Surya and reported by NOAA (blue) and KSOv (red).}
   \label{fig:diffPeak}
\end{figure}

Figures \ref{fig:diffStart} and  \ref{fig:diffPeak} show the distributions of 
the absolute differences of the flare start times and peak times, respectively, 
derived by Surya in comparison to NOAA and to KSOv. For the start and the peak 
times, the median time difference is $3$ min (1 min) with respect to NOAA 
(KSO). For $62$\% ($78$\%) of the flares detected, the derived flare start 
times lie within $\pm 5$~min with respect to the NOAA (KSOv) reports,
and for $70$\% ($83$\%) the flare peak times lie within $\pm 5$~min with 
respect to NOAA (KSOv).

\subsection{Real-time flare alerts}

When a flare reaches a certain threshold, an alert email is automatically 
generated and sent to a predefined list of users. The expected number of alerts 
is actually higher than the number of detected Surya flares  listed in Table 
\ref{tab:flares1}, since alerts are not only sent for the detection of an 
event but also in the case where the flare evolves to a higher importance class. 
This means that for a flare of importance 1, we have two alerts when the 
full flare evolution was covered: one when it reaches the level of a subflare 
of area $\ge$50~micro-hemisphere, and another one when it reaches an area of 
100~micro-hemispheres, i.e. the threshold for an importance 1 flare. 
In total, we had 14 cases (14\%) where erroneously no flare alert emails were 
issued. These include the 4 flares on July 21, 2013, the 5 flares on Oct 11, 2013 
and the first 2 flares on Oct 15, 2013, where we had an error in the automatic 
email script. In addition, no alert was sent for 3 flares that reached exactly the 
threshold area of 50 micro-hemispheres (Aug 15, 2013, 12:03 and 
and 12:49; Aug 30, 2013, 06:14). The total number of false alerts was six. 
Three of the false alerts are related to the false flare detections (indicated 
in red in Table~\ref{tab:flares1}). The other three false alerts were double 
alerts, i.e. two identical emails had been sent for one flare.

\begin{figure}
\centerline{\includegraphics[width=0.95\textwidth,clip=]{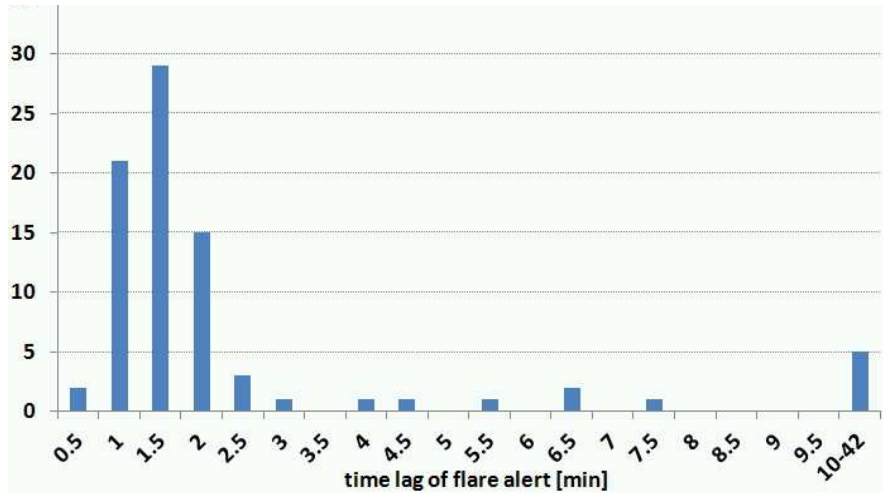}}
   \caption{Distribution of the time differences between the start of the 
   flare and the issue of the alert.}
   \label{fig:diffAlert}
\end{figure}

We also evaluated the time between the occurrence of the flare and the issue of the 
alert email. In this respect, occurrence means the time when the flare reaches 
the threshold area, i.e.\ an area of $50$ micro-hemispheres for a subflare alert, 
or an area of $100$ micro-hemispheres for an importance $1$ alert, etc. 
Figure~\ref{fig:diffAlert} shows the distribution of the time difference between 
the occurrence of the flare and the issue of the alert email, giving a median of $1.5$~min. 
In total, $89$\% of the (true) alerts were issued within $5$ minutes. 
However, we note that there are five cases where the delay is larger 
than $10$~min. These are mostly related to data gaps in the H$\alpha$ image sequences. 
The detection of a flare demands that a flare is detected in at least three observations.
However, if there is a longer data gap, say, between the second and the third image of a 
sequence, then the alert (which is issued after the flare detection in the third image) 
may occur substantially delayed with respect to the start of the flare (defined by the time of 
the first image in this sequence).

\section{Discussion and Conclusions}

The real-time H$\alpha$ data provision worked perfectly fine, with a percentage 
of $99.96$\% provisions online at the ESA SWE portal within less than 4~s of 
the observations. The automatic flare detections basically worked in all cases 
within the given criteria, i.e. within the demand that for a positive detection 
we need at least three H$\alpha$ images of quality ``good" during the event. 
In our 5-month evaluation period, about 70\% of all the H$\alpha$ images observed at KSO 
(with a regular cadence of 6 s) were classified as ``good".
We note that for visual classifications by an experienced observer also fewer 
images or images of lower quality may suffice but result in flare parameters 
with large uncertainties. This is reflected in the number of 87 flares that were 
detected by the automatic system, whereas the KSO visual reports included 19 
additional events (18 SF, one 1N).

The automatically determined flare importance and brightness classifications 
were correct in about 85\% of the events. The misclassification of 15\% is 
comparable to the $\sim$7\% (17\%) inconsistencies between the NOAA and KSOv 
flare reports for the importance (brightness) class. These differences in the 
official reports are related to different instruments, seeing conditions and 
observers that derived the parameters. The mean of the calculated heliograpic 
longitude and latitude of the flare center was consistent with the official 
flare reports within $\sim$1$^\circ$. The median of the absolute differences 
between the flare start and flare peak times of the automatic detections in 
comparison to the NOAA (KSOv) reports were 3 min (1 min). 
In $\sim$90\% of the flare alert emails that were sent, the alert was issued 
within 5 min of the flare start. However, 15\% of the expected alerts had been 
not sent. The number of false flare detections and alerts was less than 6\% 
compared to the total number of (true) alerts issued.

We note that our event set consisted mostly of subflares (69) and importance 1 
flares (18). There was one event reported as $2$N by NOAA and KSOv 
(October 10, 2013; cf. Table \ref{tab:flares1}), which was misclassified as 
$1$F by our automatic algorithm. Re-inspection of the processing of this event 
showed that the area had been correctly calculated (i.e. exceeded the threshold 
to an importance 2 flare), but due to a large data gap during the observing 
sequence, the algorithm erroneously applied a wrong time (during the flare 
decay phase) for the calculation of the importance class. 

We conclude that the automatic flare detection implemented at KSO and provided 
online at ESA's SSA SWE portal provides reliable and near real-time detection, 
classification and alerting of solar H$\alpha$ flares. The information on the 
flare timing, strength and heliographic position (which relates to the magnetic 
connectivity to Earth) that is derived in near real-time could, e.g., be connected 
to SEP models. We note that the largest challenges of the algorithm are actually 
the handling of data gaps, which are the largest source of misclassification of 
the flare class and the split-up of one flare into more than one. Further systematic 
evaluation of the system at times of higher solar activity and more frequent 
occurrence of larger flares will be valuable in order to test its ability for 
the automatic detection of the most severe space-weather effective events.

%
 \appendix
 
Table \ref{tab:flares1} lists all flares detected by Surya with an area 
$\geq$ $50$ micro-hemispheres and located within 60$^\circ$ of the solar 
disk during the period June 26 to November 30, 2013 together with the corresponding 
information from the NOAA and KSOv flare reports. Column $1$ gives the 
observation date, columns $2-4$ list the start time of the flare
(from Surya, NOAA, KSOv), columns $5-7$ the peak time, columns $8-10$ the heliographic position, 
columns $11-13$ the flare type and column $14$ the flare area as determined by Surya.
False flare detections by Surya are marked in red color.

 \begin{sidewaystable}
 \begin{footnotesize}
  \centering
  \caption{Automatically detected flares in the period June 26 to 
      November 30, 2013. Flares marked in red color are false detections.}
    \begin{tabular}{cccccccccccccc}
    \toprule
      & \multicolumn{3}{c}{\textbf{Start time}} 
      & \multicolumn{3}{c}{\textbf{Peak time}} 
      & \multicolumn{3}{c}{\textbf{Position}} 
      & \multicolumn{3}{c}{\textbf{Type}} & \textbf{Area} \\
    \midrule
      & \textbf{Surya} & \textbf{NOAA} & \textbf{KSOv} & \textbf{Surya} 
      & \textbf{NOAA} & \textbf{KSOv} & \textbf{Surya} & \textbf{NOAA} 
      & \textbf{KSOv} & \textbf{Surya} & \textbf{NOAA} & \textbf{KSOv} 
      & \textbf{Surya} \\ 
    \midrule
      \textbf{27/06} & 09:38 & B09:38 & 09:38 & 09:41 & U09:42 & 09:41 & S16E23 
        & S16E25 & S16E23 & SF    & SF & SF & 60 \\ 
    \midrule
      \textbf{30/06} & 09:11 & 09:14 &  & 09:33 & 09:16 & & S15W20 & S16W19 &  
        & SF  & SF & & 74 \\ 
    \midrule
      \textbf{04/07} & \color{red}{09:21} & & & \color{red}{09:22} & & & 
      \color{red}{S11E45} &  &  & \color{red}{SF} &  &  & \color{red}{50} \\ 
    \midrule
      \multirow{2}[1]{*}{\textbf{05/07}} & 04:26 & B05:25 & 05:01 & 05:16 & 
        U05:34 & 05:16 & S08E30 & S06E38 & S08E30 & SF & SF & SF & 87 \\
      & 06:57 & 06:58 & 06:57 & 06:59 & 07:03 & 06:59D & S09E30 & S09E33 & 
        S09E29 & SF & SF & SF & 60 \\ 
    \midrule
      \multirow{3}[0]{*}{\textbf{09/07}} & 07:38 &  & 07:38E & 07:40 &  & 07:40U 
        & S08W23 & & S13W18 & 1F    &  & 1F & 156 \\
      & \color{red}{08:32} & & & \color{red}{08:36} & & & \color{red}{S08W24} 
        &  &  & \color{red}{SF} &  &  & \color{red}{115} \\
      & 13:27 & 13:26 & 13:27 & 13:31 & 13:31 & 13:32 & S10W21 & S12W21 & 
        S10W21 & SN & 1N & SN & 82 \\ 
    \midrule
      \multirow{2}[0]{*}{\textbf{10/07}} & 06:20 & 06:21 & 06:20U & 06:32 & 
        06:43 & 06:31D & S14W13 & S15W13 & S14W13 & 1F & 1N & 1N & 177 \\
      &  \color{red}{07:07} &  &  & \color{red}{07:08} &  &  & \color{red}{S13W15} 
        &  &  & \color{red}{SF} &  &  & \color{red}{72} \\ 
    \midrule
      \textbf{16/07} & 10:11 & 10:12 & 10:10 & 10:20 & 10:16 & 10:20 & S12E03 & 
        S12E04 & S12E03 & SF & SF & SF & 96 \\ 
    \midrule
      \multirow{4}[2]{*}{\textbf{21/07}} & 06:41 & 06:43 & 06:41 & 06:44 & 06:44 
        & 06:45 & N22W07 & N23W07 & N22W08 & SF & SF & SF & 61 \\
      & 08:25 & 08:25 & 08:25 & 08:30 & 08:41 & 08:31 & S07E30 & S06E31 
        & S07E31 & 1F & 1F & 1F & 151 \\
      & 12:16 & 12:17 & 12:17 & 12:19 & 12:18 & 12:18 & N22W09 & N22W09 
        & N22W09 & SN & SF & SN & 51 \\
      & 14:13 &  & 14:11 & 14:13 &  & 14:14 & N23W11 &  & N24W11 & SF &  
        & SF & 65 \\ 
    \midrule
      \textbf{25/07} & 06:05 & 06:05 & 06:05E & 06:08 & 06:07 & 06:08 & S06W20 
        & S08W22 & S06W20 & SN & SF & SN & 59 \\ 
    \midrule
      \textbf{28/07} & 12:07 & 12:05 & 12:05 & 12:23 & 12:23 & 12:22 & S11W59 
        & S13W60 & S11W59 & SF & SF & SF & 70 \\ 
    \midrule
      \textbf{06/08} & 08:03 & 08:04 & 08:04 & 08:04 & 08:04 & 08:04 & N17W01 
        & N15W01 & N17W00 & SF & SF & SF & 65 \\ 
    \midrule
      \multirow{2}[1]{*}{\textbf{08/08}} & 05:33 & \multirow{2}[1]{*}{ 05:39} 
        & \multirow{2}[1]{*}{05:31E} & 05:34 & \multirow{2}[1]{*}{05:43} 
        & \multirow{2}[1]{*}{05:45U} & S13W10 & \multirow{2}[1]{*}{S14W08} 
        & \multirow{2}[1]{*}{S13W09} & SF & \multirow{2}[1]{*}{SF} 
        & \multirow{2}[1]{*}{SF} & 54 \\
      & 05:41 &  & & 05:42 &  &  & S13W08 &  & & SF &  &  & 60 \\ 
    \midrule
      \textbf{11/08} & 07:09 & & 07:51 & 08:04 &  & 07:57 & S20E31 & & S20E31 
        & SF & & SF & 54 \\ 
    \midrule
      \multirow{2}[1]{*}{\textbf{12/08}} & \multirow{2}[1]{*}{ 09:49} & 09:49 & 09:49 
        & \multirow{2}[1]{*}{10:43} & 09:49 & 09:53 
        & \multirow{2}[1]{*}{S21E18} & S17E20 & S21E19 
        & \multirow{2}[1]{*}{1N} & SF & SF & \multirow{2}[1]{*}{141} \\
      &  & 10:23 & 10:22 &  & 10:40 & 10:37 &  & S17E19 & S21E19 & & SN & 1N &  \\ 
    \midrule
      \multirow{3}[1]{*}{\textbf{15/08}} & 12:03 
        & \multirow{3}[1]{*}{} & 12:03 & 12:20 
        & \multirow{3}[1]{*}{} & 12:20 & S06E04 
        & \multirow{3}[1]{*}{} & S06E04 & SF 
        & \multirow{3}[1]{*}{} & SF & 50 \\
      & 12:37 &  & 12:37 & 12:44 &  & 12:45 & S06E01 &  & S06E01 & SF &  & SF & 54 \\
      & 12:49 &  &  & 12:49 & &  & S06E01 &  &  & SF  &  &  & 50 \\
    \bottomrule
    \end{tabular}%
  \label{tab:flares1}%

  \end{footnotesize}
\end{sidewaystable}

\begin{sidewaystable}
\begin{footnotesize}
  \caption{Table \ref{tab:flares1} continued.}
    \begin{tabular}{cccccccccccccc}
    \toprule
      & \multicolumn{3}{c}{\textbf{Start time}} 
      & \multicolumn{3}{c}{\textbf{Peak time}} 
      & \multicolumn{3}{c}{\textbf{Position}} 
      & \multicolumn{3}{c}{\textbf{Type}} 
      & \textbf{Area} \\
    \midrule
      & \textbf{Surya} & \textbf{NOAA} & \textbf{KSOv} & \textbf{Surya} 
      & \textbf{NOAA} & \textbf{KSOv} & \textbf{Surya} & \textbf{NOAA} 
      & \textbf{KSOv} & \textbf{Surya} & \textbf{NOAA} & \textbf{KSOv} 
      & \textbf{Surya} \\ 
    \midrule
      \multirow{2}[2]{*}{\textbf{17/08}} & \multirow{2}[2]{*}{10:18} 
        & 10:11 & 10:09 & \multirow{2}[2]{*}{10:55} & 10:18 & 10:19 
        & \multirow{2}[2]{*}{S05W28} & S08W24 & S07W23 & \multirow{2}[2]{*}{SN} 
        & SF & SN & \multirow{2}[2]{*}{61} \\
      & & 10:55 & 10:53 & & 10:56 & 10:56 & & S07W25 & S05W28 &  & SF & SN &  \\ 
    \midrule
      \textbf{19/08} & 05:42 & & 05:43 & 05:48 & & 05:46 & N15W16 & & N15W17 
        & SF & & SF & 63 \\ 
    \midrule
      \multirow{4}[4]{*}{\textbf{21/08}} & 07:13 & 07:17 & 07:17 & 07:18 & 07:17 
        & 07:18 & S13W38 & S15W36 & S13W38 & SF & SF & SF & 78 \\
      & 07:30 & 07:29 & 07:29 & 07:30 & 07:42 & 07:37 & N16E57 
        & N14E55 & N16E57 & SF & SF & SF & 98 \\
      & \multirow{2}[2]{*}{08:33} & B09:06 & \multirow{2}[2]{*}{09:06} 
        & \multirow{2}[2]{*}{09:11} & U09:11 & \multirow{2}[2]{*}{09:10} 
        & \multirow{2}[2]{*}{S13W38} & S15W37 & \multirow{2}[2]{*}{S13W38} 
        & \multirow{2}[2]{*}{SF} & SF & \multirow{2}[2]{*}{SF} 
        & \multirow{2}[2]{*}{75} \\ 
      & & 09:09 & & & 09:09 & & & S14W40 & & & SF & & \\ 
    \midrule
      \textbf{22/08} & 11:45 & B11:49 & 11:46 & 11:48 & U11:58 & 11:47 & N15E37 
        & N17E38 & N14E35 & SF    & SF & SF & 65 \\ 
    \midrule
      \textbf{23/08} & 05:18 & & 05:17E & 05:19 &  & 05:18 & N12W53 &  
        & N12W52 & SF &   & SN & 69 \\ 
    \midrule
      \textbf{30/08} & 06:14 & & & 06:14 &  &  & N16E44 &  &  & SF &   
        &  & 50 \\ 
    \midrule
      \multirow{12}[2]{*}{\textbf{04/09}} & 05:32 & \multirow{4}[4]{*}{05:34}  
        & \multirow{2}[2]{*}{05:32} & 05:40 &  \multirow{4}[4]{*}{05:39}  
        & \multirow{2}[2]{*}{05:40} & S16W35 &  \multirow{4}[4]{*}{S19W35} 
        & \multirow{2}[2]{*}{S16W35} & SB &  \multirow{4}[4]{*}{SF} 
        & \multirow{2}[2]{*}{SB} & 82 \\ 
        & 06:39 & & & 06:47 & & & S16W36 & & & SF & & & 59  \\
      & 06:56 &  & \multirow{4}[4]{*}{06:57} & 07:01 &  
        & \multirow{4}[4]{*}{07:23} & S16W36 &  & \multirow{4}[4]{*}{S16W36}  
        & SF &  & \multirow{4}[4]{*}{SF} & 72 \\  
      & 07:11 &  &  & 07:12 &  &  & S16W37 &  &  & SF &  &  & 77 \\
      & 07:18 & 07:21 &  & 07:20 & 07:26 &  & S16W37 & S16W38 &  & SF & SF &  & 88 \\
      & 07:50 &  &  & 07:51 &  &  & S16W37 &  &  & SF &  &  & 59 \\
      & 07:55 &  & \multirow{2}[2]{*}{07:59}  & 07:59 &  & \multirow{2}[2]{*}{08:03}  
        & S16W36 &  & \multirow{2}[2]{*}{S16W36} & SF &  & \multirow{2}[2]{*}{SF} & 66 \\  
      & 08:02 &  &  & 08:03 &  &  & S16W36 &  &  & SF &  &  & 69 \\
      & 08:21 & 08:18 & 08:17 & 08:42 & 08:39 & 08:40 & S16W36 & S18W37 
        & S16W37 & SF & SF & 1N & 79 \\
      & 10:16 & 09:59 & 09:59 & 10:16 & 10:18 & 10:16 & S16W37 & S18W36 & S16W37 
        & SF & SF & SF & 66 \\
      & \multirow{2}[2]{*}{11:42} & B11:45 & \multirow{2}[2]{*}{11:50U} 
        & \multirow{2}[2]{*}{12:30} & U11:45 & \multirow{2}[2]{*}{12:19} 
        & \multirow{2}[2]{*}{S16W39} & S18W38 & \multirow{2}[2]{*}{S16W39} 
        & \multirow{2}[2]{*}{SF} & SF & \multirow{2}[2]{*}{SF} 
        & \multirow{2}[2]{*}{54} \\
      &  & 12:08 &  &  & 12:14 &  & & S18W38 &  &  & SF &  &  \\ 
    \midrule
      \multirow{4}[4]{*}{\textbf{11/10}} & 12:19	& & 12:19E & 12:21 & 
        & 12:20U & S24E40 & & S24E43 & SF & & SF & 71 \\
      & 14:16 & 14:34 & 14:33 & 14:38 & 14:46 & 14:47 & S09E13 & S09E14 
        & S09E13 & 1N & SF & SF & 130 \\
      & 15:06 & & & 15:08 & & & S11E14 & & & SF & & & 91 \\
      & 15:25 & & & 15:26 & & & S09E12 & & & SF & & & 90 \\    
    \bottomrule
    \end{tabular}%
  \label{tab:flares2}%
  \end{footnotesize}
\end{sidewaystable}

\begin{sidewaystable}
\begin{footnotesize}
  \caption{Table \ref{tab:flares1} continued.}
    \begin{tabular}{cccccccccccccc}
    \toprule
       & \multicolumn{3}{c}{\textbf{Start time}} 
       & \multicolumn{3}{c}{\textbf{Peak time}} 
       & \multicolumn{3}{c}{\textbf{Position}} 
       & \multicolumn{3}{c}{\textbf{Type}} & \textbf{Area} \\
    \midrule
       & \textbf{Surya} & \textbf{NOAA} & \textbf{KSOv} & \textbf{Surya} 
       & \textbf{NOAA} & \textbf{KSOv} & \textbf{Surya} & \textbf{NOAA} 
       & \textbf{KSOv} & \textbf{Surya} & \textbf{NOAA} & \textbf{KSOv} 
       & \textbf{Surya} \\ 
    \midrule
      \multirow{4}[4]{*}{\textbf{15/10}} & 06:50 & & 06:49 & 07:02 &  & 07:01 
        & S21W14 &  & S22W15  & SF &  & SF & 76 \\ 
      & \multirow{2}[2]{*}{08:02} & 08:14 & 08:01E 
        & \multirow{2}[2]{*}{08:39} & 08:15 & 08:01E 
        & \multirow{2}[2]{*}{S21W13} & S22W14 & S21W13  
        & \multirow{2}[2]{*}{1N} & SF & SF & \multirow{2}[2]{*}{235} \\   
      &  & 08:33 & 08:23U &  & 08:39 & 08:39 &  & S22W13 & S21W13  
        &  & SN & 1B &  \\ 
      & 08:57 & 09:09 & 09:00 & 09:14 & 09:15 & 09:14 & S10W35 & S10W36 
        & S10W35  & SF & SF & SF & 80 \\
    \midrule
      \multirow{3}[3]{*}{\textbf{16/10}} & 08:51 & 09:12 & 09:12 & 09:20 
        & U09:20 & 09:20 & S09W42 & S11W41 & S09W42  & SF & SF & SF & 94 \\ 
      & 13:55 &  & 13:55 & 13:58 &  & 13:58 & S09W44 &  & S10W44  & SF 
        &  & SF & 72 \\
      & 14:27 & 14:27 & 14:26 & 14:31 & 14:31 & 14:32 & S21W30 & S23W29 
        & S21W30  & 1N & 1N & 1B & 185 \\ 
    \midrule 
      \multirow{2}[2]{*}{\textbf{17/10}} & 10:21 & 10:27 & 10:26 & 10:29 
        & 10:29 & 10:29 & S09W56 & S11W56 & S09W56  & SF & SN & SN & 74 \\ 
      & 11:15 & 11:48 & 11:47 & 13:15 & U12:03 & 12:04 & S09W58 
        & S11W56 & S10W54 & SF & SF & SF & 61 \\  
    \midrule 
      \multirow{5}[3]{*}{\textbf{20/10}} & 08:38 & 08:36 & 08:36 & 08:42 & 08:39 
        & 08:41 & N22W32 & N22W33 & N22W32  & 1N & SF & 1N & 133 \\ 
      & 09:39 & 09:40 & 09:40 & 09:57 & 09:43 & 09:50 & N12E43 & N11E40 
        & N12E42 & 1F & SF & SF & 135 \\
      & 10:01 &  &  & 10:22 &  &  & N09E37 &  &  & 1F &  &  & 136 \\
      & 11:54 & 11:58 & 11:55 & 12:48 & U12:16 & 12:19 & N11E41 & N12E39 
        & N11E41 & 1F & SF & SF & 167 \\
      & 12:41 & 12:42 & 12:40 & 12:48 & U12:45 & 12:46 & N22W34 & N22W35 
        & N22W34 & SF & SF & SF & 75 \\
    \midrule   
      \multirow{3}[3]{*}{\textbf{22/10}} & 07:35 &  & 07:35 & 07:41 &  & 07:42 
        & N09E16 &  & N09E15  & SF & & SF & 73 \\ 
      & 08:33 &  & 08:29U & 08:34 &  & 08:34U & N06E13 &  & N06E13 & SF &  
          & SF & 85 \\
      & 13:30 & 13:27 & 13:27 & 13:32 & 13:29 & 13:31 & N05E04 & N06E04 
        & N04E04 & SF & SF & SN & 94 \\ 
    \midrule  
      \multirow{11}[3]{*}{\textbf{26/10}} & 07:33 &  &  & 07:36 &  & & N04W38 
        &  &   & SF & &  & 90 \\   
      & 07:50 &  &  & 07:56 &  & & N04W38 &  &   & SF & &  & 58 \\
      & 07:58 &  & 07:58 & 08:04 &  & 08:03 & S10W22 &  & S10W22  & SN & 
        & SF & 74 \\  
      & 08:23 & B08:54 & 08:26 & 08:56 & U08:55 & 08:55 & S11W23 & S12W23 
        & S11W23  & 1B & SF & 1B & 127 \\ 
      & 08:47 &  & 09:18 & 09:19 &  & 09:21 & S09E55 &  & S09E54  & 1B & 
        & 1B & 139 \\ 
      & 09:01 &  & 09:23 & 09:25 &  & 09:25 & N07W48 &  & N07W49  & 1B & 
        & 1B & 123 \\
      & 09:26 &  &  & 09:32 &  & & S11W23 &  &   & SN & &  & 72 \\
      & 10:22 &  &  & 10:27 &  & & S07E57 &  &   & SF & &  & 59 \\
        & 10:32 & B10:11 & 10:30 & 11:00 & U11:08 & 11:10 & S06E58 & S05E58 
      & S06E58  & 1N & 1N & 1N & 130 \\
      & 10:54 &  & 10:54 & 11:08 &  & 11:08 & S11W24 &  & S11W24  & SF & & SF & 57 \\ 
      & 13:18 & 13:18 & 13:18 & 13:20 & 13:20 & 13:20 & S11W26 & S12W25 & S11W26 
        & SF & SF & SF & 71 \\
    \midrule                         
    \bottomrule
    \end{tabular}%
  \label{tab:flares3}%
  \end{footnotesize}
\end{sidewaystable}

\begin{sidewaystable}
\begin{footnotesize}
  \caption{Table \ref{tab:flares1} continued.}
    \begin{tabular}{cccccccccccccc}
    \toprule
      & \multicolumn{3}{c}{\textbf{Start time}} 
      & \multicolumn{3}{c}{\textbf{Peak time}} 
      & \multicolumn{3}{c}{\textbf{Position}} 
      & \multicolumn{3}{c}{\textbf{Type}} & \textbf{Area} \\
    \midrule
      & \textbf{Surya} & \textbf{NOAA} & \textbf{KSOv} & \textbf{Surya} 
      & \textbf{NOAA} & \textbf{KSOv} & \textbf{Surya} & \textbf{NOAA} 
      & \textbf{KSOv} & \textbf{Surya} & \textbf{NOAA} & \textbf{KSOv} 
      & \textbf{Surya} \\ 
    \midrule
      \multirow{3}[3]{*}{\textbf{27/10}} & 10:18 & & 10:16E & 10:19 &  & 10:18U 
        & S08E43 &  & S08E43  & SF &  & SF & 54 \\ 
      & 11:00 & & 10:59 & 11:00 &  & 11:01 & S08E43 &  & S08E43  & SF &  
        & SF & 69 \\ 
      & 12:13 & 12:33 & 12:30 & 12:39 & 12:38 & 12:39 & N08W63 & N06W63 
        & N08W63 & SF & 1F & 1F & 80 \\ 
    \midrule
      \multirow{3}[3]{*}{\textbf{28/10}} & 11:10 & 11:25& 11:28E & 11:29 
        & 11:36 & 11:28 & S06E27 & S06E30 & S07E28 & SF & SF & SF & 51 \\ 
      & 11:52 & 11:33 & 11:50E & 11:52 & 11:49 & 11:51 & S14W43 & S16W44 &
        S14W43  & 1F & 2N & 2N & 122 \\ 
      & 12:13 &  & 11:50E & 12:13 & & 11:53 & S07E28 &  & S05E31  & 1F &
        & 1N & 154 \\ 
    \midrule
      \multirow{4}[4]{*}{\textbf{01/11}} & \multirow{2}[2]{*}{09:26} & B09:51 &  
        & \multirow{2}[2]{*}{10:08} & 09:54 & 
        & \multirow{2}[2]{*}{S11E04} & S12E06 &  
        & \multirow{2}[2]{*}{SB} & SF &  & 63 \\ 
      &  & 10:05  &  &  & 10:12 & &  & S12E03 &  &  & SF &  &  \\
      & 10:50 &  &  & 10:51 &  & & S16E07 &  &  & SF &  &  & 61 \\
      & 11:08 &  &  & 11:09 & & & S17E08 &  &  & SF &  &  & 51 \\
    \midrule 
      \multirow{2}[2]{*}{\textbf{07/11}} & 10:28 & 10:28 & 10:28 & 10:29 & 10:29
        & 10:31 & N21W36 & N21W37 & N21W37 & SF & SF & SF & 67 \\ 
        & 12:16 & 12:17 & 12:17 & 12:28 & 12:28 & 12:28 & S13E23 & S13E24 &
        S13E23 & SN & SF & SN & 55 \\
    \midrule
       \multirow{1}[1]{*}{\textbf{29/11}} & 09:55 & 10:03 & 09:55 & 10:10 & 10:06
         & 10:10 & S06W23 & S05W16 & S06W23 & 1F & SF & 1F & 145 \\ 
    \midrule
    \bottomrule
    \end{tabular}%
  \label{tab:flares4}%
  \end{footnotesize}
\end{sidewaystable}

  
\begin{acks}
This study was developed within the framework of ESA Space Situational Awareness 
(SSA) Programme (SWE SN IV-2 activity). The authors thank Alexi Glover and 
Juha-Pekka Luntama for their support, constructive criticism and confidence in 
the project.
\end{acks}

\bibliographystyle{spr-mp-sola}
\tracingmacros=2
\bibliography{sniv_solphy_28102014}  

\end{article} 

\end{document}